\documentclass[fleqn,usenatbib]{mnras}

\usepackage{newtxtext,newtxmath}

\usepackage[T1]{fontenc}
\usepackage{ae,aecompl}

\usepackage{graphicx}	
\usepackage{amsmath}	
\usepackage{amssymb}	
\usepackage[dvipsnames]{xcolor}		
\usepackage{hyperref}
\usepackage{subfigure}

\def\CIVdblt{{\rm C~}\kern 0.1em{\sc iv}~$1548, 1550$}
\def\NVdblt{{\rm N~}\kern 0.1em{\sc v}~$1238, 1242$}
\def\OVIdblt{{\rm O~}\kern 0.1em{\sc vi}~$ 1031, 1037$}
\def\SiIVdblt{{\rm Si~}\kern 0.1em{\sc iv}~$1394, 1403$}
\def\NeVIII{\hbox{{\rm Ne~}\kern 0.1em{\sc viii}}}
\def\OII{\hbox{{\rm O~}\kern 0.1em{\sc ii}}}
\def\OIII{\hbox{{\rm O~}\kern 0.1em{\sc iii}}}
\def\OIV{\hbox{{\rm O~}\kern 0.1em{\sc iv}}}
\def\OV{\hbox{{\rm O~}\kern 0.1em{\sc v}}}
\def\OVI{\hbox{{\rm O~}\kern 0.1em{\sc vi}}}
\def\OVII{\hbox{{\rm O~}\kern 0.1em{\sc vii}}}
\def\OVIII{\hbox{{\rm O~}\kern 0.1em{\sc viii}}}
\def\NII{\hbox{{\rm N~}\kern 0.1em{\sc ii}}}
\def\NIII{\hbox{{\rm N~}\kern 0.1em{\sc iii}}}
\def\NIV{\hbox{{\rm N~}\kern 0.1em{\sc iv}}}
\def\NV{\hbox{{\rm N~}\kern 0.1em{\sc v}}}
\def\NVII{\hbox{{\rm N~}\kern 0.1em{\sc vii}}}
\def\CII{\hbox{{\rm C~}\kern 0.1em{\sc ii}}}
\def\CIII{\hbox{{\rm C~}\kern 0.1em{\sc iii}}}
\def\SiIII{\hbox{{\rm Si~}\kern 0.1em{\sc iii}}}
\def\SIV{\hbox{{\rm S~}\kern 0.1em{\sc iv}}}
\def\SV{\hbox{{\rm S~}\kern 0.1em{\sc v}}}
\def\SVI{\hbox{{\rm S~}\kern 0.1em{\sc vi}}}
\def\SiII{\hbox{{\rm Si~}\kern 0.1em{\sc ii}}}
\def\SiIV{\hbox{{\rm Si~}\kern 0.1em{\sc iv}}}
\def\SiI{\hbox{{\rm Si~}\kern 0.1em{\sc i}}}
\def\PII{\hbox{{\rm P~}\kern 0.1em{\sc ii}}}
\def\AlII{\hbox{{\rm Al~}\kern 0.1em{\sc ii}}}
\def\AlIII{\hbox{{\rm Al~}\kern 0.1em{\sc iii}}}
\def\MgII{\hbox{{\rm Mg~}\kern 0.1em{\sc ii}}}
\def\FeII{\hbox{{\rm Fe~}\kern 0.1em{\sc ii}}}
\def\CaI{\hbox{{\rm Ca~}\kern 0.1em{\sc i}}}
\def\CaII{\hbox{{\rm Ca~}\kern 0.1em{\sc ii}}}
\def\CrII{\hbox{{\rm Cr~}\kern 0.1em{\sc ii}}}
\def\CII{\hbox{{\rm C~}\kern 0.1em{\sc ii}}}
\def\CIII{\hbox{{\rm C~}\kern 0.1em{\sc iii}}}
\def\CIV{\hbox{{\rm C~}\kern 0.1em{\sc iv}}}
\def\CV{\hbox{{\rm C}\kern 0.1em{\sc v}}}
\def\H{\hbox{{\rm H~}}}
\def\HI{\hbox{{\rm H~}\kern 0.1em{\sc i}}}
\def\HII{\hbox{{\rm H~}\kern 0.1em{\sc ii}}}
\def\Lya{\hbox{{\rm Ly}\kern 0.1em$\alpha$}}
\def\Lyb{\hbox{{\rm Ly}\kern 0.1em$\beta$}}
\def\Lyg{\hbox{{\rm Ly}\kern 0.1em$\gamma$}}
\def\Lyth{\hbox{{\rm Ly}\kern 0.1em$\theta$}}
\def\Lyfive{\hbox{{\rm Ly}\kern 0.1em$5$}}
\def\Lysix{\hbox{{\rm Ly}\kern 0.1em$6$}}
\def\Lyseven{\hbox{{\rm Ly}\kern 0.1em$7$}}
\def\Lyeight{\hbox{{\rm Ly}\kern 0.1em$8$}}
\def\Lynine{\hbox{{\rm Ly}\kern 0.1em$9$}}
\def\Lyten{\hbox{{\rm Ly}\kern 0.1em$10$}}
\def\kms{\hbox{km~s$^{-1}$}}
\def\cmsq{\hbox{cm$^{-2}$}}
\def\cc{\hbox{cm$^{-3}$}}
\usepackage{soul}

\usepackage{tabularx,ragged2e}		
\usepackage{booktabs,array,times}

\title[Solar-Metallicity Gas in $z \sim 0.12$ Galaxy]{Solar-Metallicity Gas in the Extended Halo of a Galaxy at $\MakeLowercase{z} \sim 0.12$}

\author[Pradeep et al.]{Jayadev Pradeep,$^{1}$\thanks{E-mail: jayadev\_pradeep@yahoo.com} Sriram Sankar,$^{2}$ T. M. Umasree,$^{3}$ Anand Narayanan,$^{1}$\thanks{E-mail: anand@iist.ac.in} 
\newauthor Vikram Khaire,$^{4}$ Matthew Gebhardt,$^{5}$ Sameer,$^{5}$  and Jane C. Charlton$^{5}$\\
$^{1}$Department of Earth and Space Sciences, Indian Institute of Space Science \& Technology, Thiruvananthapuram 695547, Kerala, INDIA\\
$^{2}$Department of Mechanical Engineering, Federal Institute of Science And Technology, Ernakulam 683577, Kerala, INDIA\\
$^{3}$Department of Physics, National Institute of Technology Karnataka, Mangaluru 575025, Karnataka, INDIA\\
$^{4}$Department of Physics, University of California, Santa Barbara 93106, California, USA\\
$^{5}$Department of Astronomy \& Astrophysics, The Pennsylvania State University, 525 Davey Laboratory
University Park, PA, 16802, USA\\
}

\date{Accepted 2020 January 14. Received 2019 December 22; in original form 2019 November 07}

\pubyear{2019}

\begin{document}
\label{firstpage}
\pagerange{\pageref{firstpage}--\pageref{lastpage}}
\maketitle

\begin{abstract}
We present the detection and analysis of a weak low-ionization absorber at $z = 0.12122$ along the blazar sightline PG~$1424+240$, using spectroscopic data from both $HST$/COS and STIS. The absorber is a weak {\MgII} analogue, with incidence of weak {\CII} and {\SiII}, along with multi-component {\CIV} and {\OVI}. The low ions are tracing a dense ($n_{\H} \sim  10^{-3}$~{\cc}) parsec scale cloud of solar or higher metallicity. The kinematically coincident higher ions are either from a more diffuse ($n_{\H} \sim 10^{-5} - 10^{-4}$~{\cc}) photoionized phase of kiloparsec scale dimensions, or are tracing a warm (T $\sim 2 \times 10^{5}$~K) collisionally ionized transition temperature plasma layer. The absorber resides in a galaxy overdense region, with 18 luminous ($>$ L*) galaxies within a projected radius of 5 Mpc and $750$~{\kms} of the absorber. The multi-phase properties, high metallicity and proximity to a $1.4L^*$ galaxy, at $\rho \sim 200$~kpc and $|\Delta v| = 11$~{\kms} separation, favors the possibility of the absorption tracing circumgalactic gas. The absorber serves as an example of weak {\MgII}-{\OVI} systems as a means to study multiphase high velocity clouds in external galaxies.


\end{abstract}



\section{Introduction}

Much of our knowledge of the gaseous halos of galaxies has come from quasar absorption line observations (see reviews by \citealt{putman_gaseous_2012}, \citealt{tumlinson_circumgalactic_2017}). The extended gaseous environment is best studied in the case of the Milky Way through observations of intermediate and high velocity clouds (IVCs, and HVCs), which are multiphase gas clouds moving through the hot halo of the Galaxy (e.g., \citealt{wakker_high-velocity_1997}, \citealt{putman_halpha_2003}, \citealt{savage_extension_2009}, \citealt{richter_hst/cos_2017}, \citealt{bish_galactic_2019}). Along with being baryon-rich, these clouds that occupy the halo are also tracers of inflow and outflow processes that transport material between the star-forming disk and the intergalactic medium outside of the virial bounds of the galaxy. 

Measuring the chemical abundances, ionization conditions, kinematics, and spatial distribution of these circumgalactic clouds is the most effective way to identify the gas accretion and feedback processes operating through the CGM of galaxies. For the Milky Way, such detailed insights through multiple sightline observations through the halo have lead to the understanding of the circumgalactic clouds having a variety of origins. The metal-rich clouds, most of which are at intermediate velocities of $30$~{\kms} $< \Delta v < 90$~{\kms} relative to the LSR, are presumed to be interstellar gas expelled from the disk as galactic fountains powered by supernova events (\citealt{shapiro_consequences_1976},  \citealt{bregman_galactic_1980}, \citealt{norman_disk-halo_1989}, \citealt{houck_low-temperature_1990}, \citealt{richter_molecular_2001}, \citealt{sarkar_clues_2017}). The spatially extended streams of sub-solar and higher metallicity gas in the halo are known to be gas tidally displaced from dwarf satellites (\citealt{lu_metallicity_1998}, \citealt{sembach_fuse_2001}), whereas the metal poor pristine material is thought to be gas accreted from the intergalactic medium (\citealt{wakker_accretion_1999}, \citealt{sembach_deuterium--hydrogen_2004}). Accurate estimation of chemical abundances for a wide range of elements has been a crucial piece of information in establishing these origins for the population of IVCs and HVCs (\citealt{wakker_accretion_1999}, \citealt{richter_fuse_2001}, \citealt{richter_diversity_2001}, \citealt{collins_survey_2003}, \citealt{tripp_complex_2003}, \citealt{sembach_deuterium--hydrogen_2004}, \citealt{fox_multiphase_2005}, \citealt{fox_exploring_2010}).  

Ionization models of gaseous halos have also given deep insights on the multiphase complexity of the Circumgalactic Medium (CGM). Once again, in the case of the Milky Way, such models constrained by a mix of low ({\CII}, {\MgII}, {\SiII}, and {\FeII}), intermediate ({\CIII}, {\OIII}, {\SiIII}, and {\OIV}) and high ions ({\CIV}, {\SiIV}, {\NV} and {\OVI}), have offered information on the temperature, pressure and densities in these clouds, and have also indirectly established the presence of a fully ionized and hot ($T \sim 10^6$~K) corona enveloping the Galaxy, which is otherwise hard to detect (e.g., \citealt{savage_observational_1979}, \citealt{savage_ultraviolet_1981}, \citealt{sembach_observations_1992}, dn{\citealt{savage_absorption_1997}}, \citealt{wakker_far_2003}, \citealt{sembach_highly_2003}, \citealt{fox_origins_2003}). Identifying intervening absorbers that are close analogues of Milky Way HVCs is an important step in capturing baryon inflows, outflows and recycling in external galaxies (see \citealt{schulman_ugc_1997}, \citealt{schulman_vla_1997},  \citealt{schulman_high-velocity_1996} and \citealt{swaters_h_1997} for examples). The weak {\MgII} class of quasar absorption systems at $z > 0$ have previously been proposed as proxies of high velocity gas in the CGM of other galaxies (\citealt{rigby_population_2002}, \citealt{milutinovic_nature_2006}, \citealt{narayanan_chemical_2008}, \citealt{richter_population_2009}, \citealt{muzahid_cos-weak:_2018}). 

The weak low ionization absorbers, traditionally characterized based on the rest-frame equivalent width of the {\MgII}~$2796$  transition being $W_{2796} < 0.3$~{\AA}, are unique in many ways. To begin with, unlike the strong {\MgII} systems which are statistically consistent with being Lyman limit systems tracing the disks of luminous ($> 0.1L*$) galaxies (\citealt{churchill_c_1999}, \citealt{kacprzak_halo_2008}, \citealt{gauthier_clustering_2009}, \citealt{bordoloi_radial_2011},  \citealt{churchill_definitive_2013}), the weak absorbers are found to be optically thin in neutral hydrogen with metal lines that are unsaturated (e.g. \citealt{churchill_c_1999}). The metallicity of the low ionization gas, where the {\MgII} absorption arises, typically has values greater than one-tenth solar. In many cases the best constraints are as high as ten times solar (\citealt{rigby_weak_2001}, \citealt{charlton_high-resolution_2003}, \citealt{misawa_census_2007}, \citealt{narayanan_chemical_2008}). The high metallicities imply that these absorbers should be associated with galaxies or more generally environments that are enriched by feedback from star formation. 

In the handful of studies that have explored the association of weak {\MgII} absorbers with galaxies, one or more luminous galaxies ($\gtrsim 0.05L^*$) have been found within $\sim 100$~kpc of impact parameter of the absorber (\citealt{churchill_mgii_2005}, \citealt{nielsen_magiicat_2013-1}), although the association with galaxies is not as firmly established as it is for strong {\MgII} systems. Based on a census of weak absorbers at $z < 0.3$ using $HST$/COS data (COS-WEAK survey), \citet{muzahid_cos-weak:_2018} estimated a covering fraction of $\gtrsim 30$\% for these absorbers in the CGM of galaxies brighter than $0.001L^*$, with a possible increase in covering fraction around higher luminosity galaxies (\citealt{nielsen_magiicat_2013-1}). Interestingly, this is comparable to the covering fraction of neutral HVCs as traced using {\CaII} in the extended halo of the Milky Way (\citealt{ben_bekhti_ca_2008}, \citealt{ben_bekhti_absorption-selected_2012}), pointing to the weak {\MgII} absorbers as higher redshift counterparts of Milky Way HVCs. The identification of a population of low ionization clouds in the Milky Way halo by \citet{richter_population_2009} with $N(\HI) < 10^{18}$~{\cmsq} (and therefore undetected in 21 cm emission), with associated weak absorption from {\CII}, {\MgII}, {\SiII}, and {\CaII} further supports this idea. 

A way to examine this further is to compare the ionization structure in weak {\MgII} absorbers with Milky Way HVCs. 
Galactic HVCs are found to have a multiphase structure with a dense and parsec scale neutral gas phase, as well as a highly ionized diffuse kiloparsec scale component detected through {\CIVdblt} and {\OVIdblt} absorptions in the spectra of UV-bright background sources (\citealt{sembach_far_2000}, \citealt{murphy_far_2000}, \citealt{sembach_highly_2003}, \citealt{wakker_far_2003}, \citealt{collins_highly_2004}, \citealt{fox_highly_2004}, \citealt{collins_highly_2005}, \citealt{ganguly_highly_2005}, \citealt{fox_multiphase_2005}, \citealt{fox_survey_2006}). Whereas the low ionization lines are found to be consistent with an origin in photoionized gas with $T \sim 10^4$~K, the high ions ({\CIV}, and {\OVI}) are best understood to be produced through collisional ionization in gas with warm-hot temperatures of $T \gtrsim 10^5$~K. 
\citet{Michael_Shull_2009} used the widespread detection of Si absorption in the low Milky Way halo to arrive at high infall rates ($1$~M$_{\odot}~\mathrm{yr}^{-1}$) for the Galactic HVCs and IVCs. Through pure photoionization models they inferred metallicities of less than a tenth of solar for such gas, but highlighted the large errors in the derived abundances and the possibility of unraveling hot collisionally ionized phases in such extraplanar clouds by bringing in absorption from other elements such as carbon. Such collisionally ionized gas can occur in conductive interface and/or turbulent mixing layers that form at the boundary of neutral high velocity clouds and the ambient hot coronal envelope of the Galaxy (\citealt{borkowski_radiative_1990}, \citealt{fox_highly_2004}, \citealt{fox_multiphase_2005}, \citealt{kwak_si_2015}). If some of the weak {\MgII} absorbers are high redshift HVC analogs, then they should exhibit a similar multiphase structure while tracing the CGM of the galaxy coincident with the absorption. 

In this work, we present an intervening weak low ionization metal line absorber at $z = 0.12122$ in the line of sight towards the blazar PG~$1424+240$, using spectroscopic data from both the COS and STIS instruments aboard the $HST$. This weak {\MgII} analog (based on the weak {\CII} and {\SiII} detections) shows strong multi-component {\CIV} and {\OVI} similar to multiphase HVCs. The absorber was featured in the COS-WEAK survey of \citet{muzahid_cos-weak:_2018}, where based on modelling of just the low ionization phase, solar or supersolar metallicity was suggested. In addition to examining this conclusion, we also investigate the properties of the high ionization gas in this absorber through photoionization, collisional ionization, and hybrid models that incorporate both photoionization and collisional ionization with constraints provided by a wide range of elements. We also incorporate information on galaxies in the absorber's vicinity, and across a wider field using data from the SDSS. 

In Section~\ref{section:data}, we describe the $HST$/COS and STIS archival data that have been used for this work. Section~\ref{section:analysis} presents measurements of the lines associated with the $z = 0.12122$ absorber. The photoionization and hybrid models are discussed in Section~\ref{section:modelling}. In Section~\ref{section:galaxies} we present results from a search for SDSS galaxies in the environment centered on the absorber. A summary of the key results of the work and our conclusions on the astrophysical origin of the absorber are given in  Section~\ref{section:summary}. Throughout this work we have adopted a flat $\Lambda$CDM cosmology with H$_0$ = 69.6 km s$^{-1}$ Mpc$^{-1}$, $\Omega_{M}$ = 0.286, and $\Omega_{\Lambda}$ = 0.714 (\citealt{wright_cosmology_2006}, \citealt{bennett_1_2014}).

\section{$HST$ Observations}\label{section:data}

PG~$1424+240$, with an emission redshift of $z = 0.6010$ $\pm$ $0.003$ \citep{rovero_bl-lacertae_2016}, is one of the most distant blazars known. The target was observed by $HST$/COS in $2012$ as part of a program (ID.~$12612$) to study weak intergalactic absorption towards blazars (PI. John Stocke). The COS spectral integrations were done with G130M and G160M gratings for $3.75$~ks, and $7.92$~ks respectively. We used the coadded version of the COS spectrum from the $HST$ Spectroscopic Legacy Archive \footnote{https://archive.stsci.edu/hst/spectral\_legacy/} (\citealt{peeples_hubble_2017}) for our analysis. The coadded spectrum retrieved from the archive had $6$ pixels per $17$~{\kms} resolution element of COS, which was rebinned to the Nyquist sampling rate. The rebinned spectrum carries a $S/N \sim 13 - 25$ per resolution element over the wavelength range of $1141 - 1790$~{\AA}, with the best $S/N$ between $1250 - 1550$~{\AA} where the cumulative exposure time was highest. The blazar was also observed in $2014$ with the STIS E230M grating (ID. $3288$, PI. Amy Furniss). The STIS separate exposures were downloaded from the MAST archive, and combined using the \textsc{COADSTIS} routine  \citep{cooksey_characterizing_2008}. The final STIS spectrum, with a total integration time of $9.89$~ks, spans the wavelength range $1606 - 2365$~{\AA} at resolution of FWHM $\sim 10$~{\kms}. The $S/N$ of the STIS data is significantly lesser than expected, a possible explanation for which is that the source could have gone into a low state when the $HST$ observations were carried out.  
 
\section{DATA ANALYSIS}\label{section:analysis}

\begin{figure*}
\centering
\includegraphics[trim=0cm 0cm 0cm 0cm,width=0.95\textwidth,height=0.93\textheight]{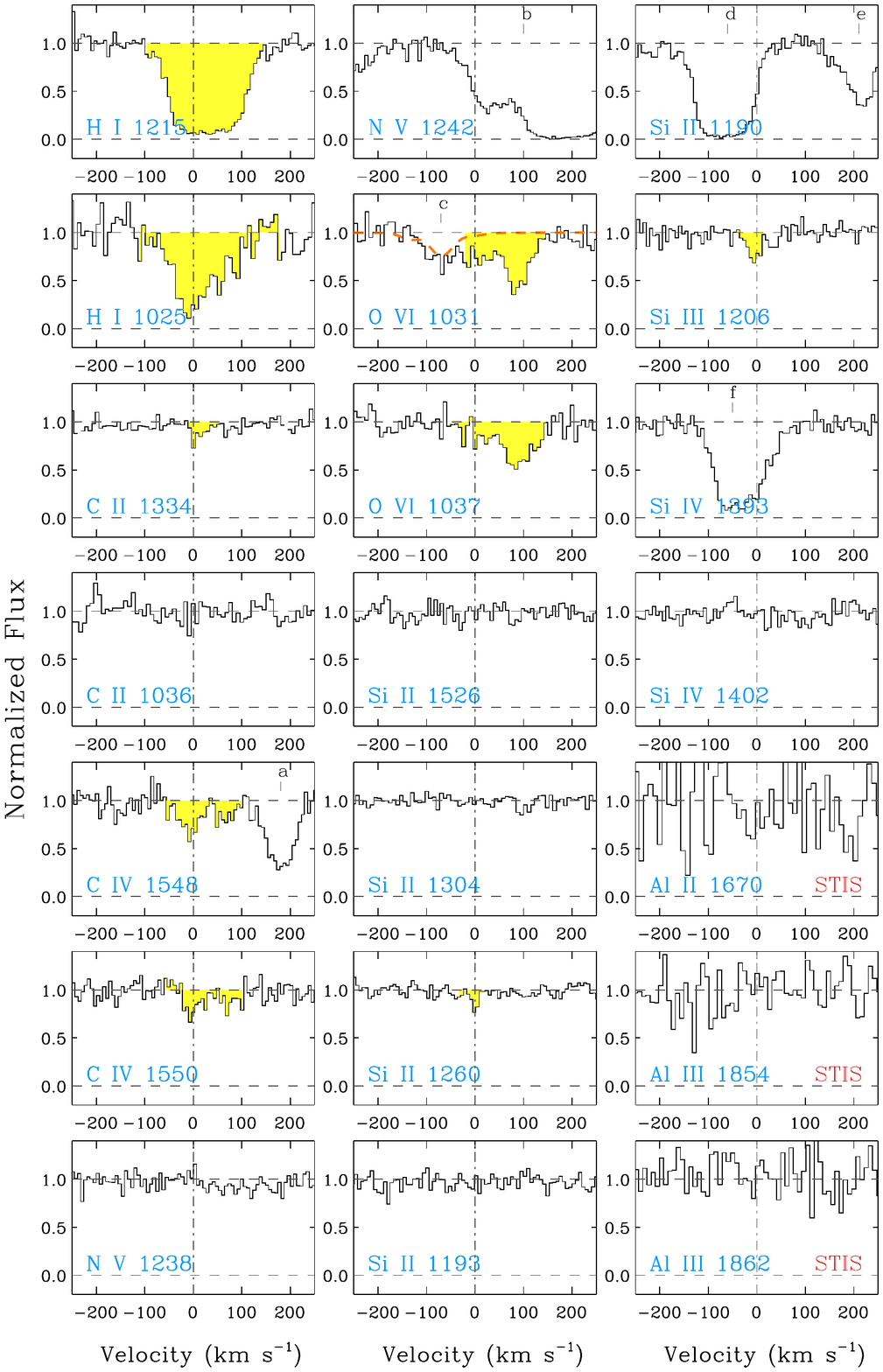}
\caption{\scriptsize{Continuum normalized spectral regions of constraining lines associated with the $z = 0.12122$ absorber towards PG~$1424+240$. Here $v = 0$~{\kms} corresponds to the centroid of the {\SiII}~$1260$ line. The highly oversampled COS spectrum has been rebinned to two pixels per $17$~{\kms} resolution element. The noisy STIS spectrum has been rebinned to COS resolution for display purposes. The relevant absorption features are indicated by the filled histogram. The aluminum lines, which are non-detections, are covered by low $S/N$ STIS E230M grating observations. The interloping lines in each panel are labeled and identified as (a) {\Lya} at $z = 0.428$, (b) Galactic {\SiIV}~$1394$ at $v \sim +30$~{\kms} and {\Lya} at $z = 0.147$ for which there are corresponding {\Lyb}, {\OVIdblt}, and {\SiIII}~$1207$ features, (c) {\Lyb} at $z = 0.128$ confirmed by corresponding {\Lya}, (d) and (e) Galactic {\CII}~$1335$ at $v = -70$~{\kms} and {\CII}$^*~1335$ at $v \sim +230$~{\kms} respectively, and (f) is likely to be {\Lya} at $z = 0.285$.} }  
\label{fig1} 
\end{figure*}

The $z = 0.12122$ absorber has {\HI}, {\CII}, {\SiII}, {\SiIII}, {\CIV}, and {\OVI} detected with $> 3\sigma$ significance. The adopted redshift is based on the centroid of the narrow {\SiII}~$1260$ line. The absorber is sufficiently displaced from the background AGN 
such that it is safe to assume it is intervening rather than intrinsic to the quasar. The continuum normalized velocity plot is shown in Figure~\ref{fig1}. The {\CIVdblt} lines clearly suggest absorption from at least two different components. A simultaneous Voigt profile fit to the {\CIV} doublets decompose the absorption into components at $v \sim -6$~{\kms} and $v \sim +63$~{\kms} (we refer to these as the \textit{central} and the \textit{offset} components). The Voigt profile fit results are listed in Table~\ref{tab1}. The fitting models were computed using the \citet{fitzpatrick_composition_1997} routine after convolving the model profiles with the appropriate COS line spread functions\footnote{http://www.stsci.edu/hst/cos/performance/spectral\_resolution/}. We refer to these two components as clouds since they are kinematically distinct, and thus also spatially distinct with possible differences in metallicity, density and temperature. 

Coincident with the central component is absorption in {\CII}, {\SiII} and {\SiIII} at a significance of $\geq 3\sigma$. The low ionization {\CII} and {\SiII} lines are very weak and only detected for this central cloud. Although the combined wavelength range of the G130M and G160M gratings does not offer coverage of {\MgII} absorption lines for this absorber, lower ionizations lines such as {\CII} and {\SiII} can be used as proxies for tracing low-ionization gas because of their similar ionization potentials (\citealt{narayanan_survey_2005}, \citealt{muzahid_cos-weak:_2018}). The rest-frame equivalent widths of $W_r(\CII~1335) = 16~{\pm}~5$~m{\AA} and $W_r(\SiII~1260) = 15~{\pm}~4$~m{\AA} make this absorber a weak {\MgII} analog \citep{narayanan_survey_2005}. The equivalent widths for all transitions, in the rest-frame of the absorber, are given in Table~\ref{tab2}. For all the lines, we have computed the integrated column densities using the AOD method developed by \citet{savage_analysis_1991}, as reported in Table \ref{tab2}. While in the case of relatively strong lines such as {\CIVdblt} and {\OVIdblt}, the apparent optical depth (AOD) derived column densities are in agreement with the column densities derived from profile-fitting, for the much weaker detections ({\SiII}, {\SiIII} and {\CII}), the profile-fitted values are $\sim 0.2$~dex higher than the column densities obtained by integrating the AOD. The difference, which is within the cumulative uncertainty between the two measurements, possibly stems from the low ionization lines being very weak. In the ionization models of Section \ref{section:modelling}, we have adopted the AOD-derived column density values for {\SiII}, {\SiIII} and {\CII}, as the AOD measurements are more conservative, taking into account the uncertainty in continuum placement in addition to the statistical uncertainty in flux values. The C and Si abundances would come out as $\sim 0.2$~dex higher if the profile-fit column densities for their respective low ions are used. The {\CII} absorption feature appears slightly broader in comparison to {\SiII} and {\SiIII}, which must be arising from the same phase. We attempted profile-fitting {\CII} by fixing the line centroid and $b$-parameter of {\CII} to the same values as {\SiII}, which resulted in a column density of $\log N(\CII) = 13.22~{\pm}~0.11$, which again is within the 1-sigma uncertainty of the free-fit measurement. The [C/H] abundance derived from the AOD column density is therefore, at worst, a lower limit.

The {\Lya} line falls in a relatively higher $S/N$ region of the spectrum, but is strongly saturated.  The {\HI} absorption spans the same velocity range as {\CIV} and {\OVI}. A large kinematic spread for the high ions with a correspondingly broad {\HI} in contrast with the narrow widths of low ions is characteristic of many weak absorbers and has been the basis for their proposed filamentary or sheet-like geometry (\citealt{milutinovic_nature_2006}). For the {\HI} component information we rely on {\Lyb} where the saturation is mild or absent at velocities away from the core absorption. Driven by the two component kinematics of {\CIV}, we applied simultaneous fits to the {\Lya} and {\Lyb} lines with two components. In such a free fit, the routine recovered {\HI} absorption at velocities close to the components seen in {\CIV}. The low ionization gas also contributes to the bluer component through {\CII}, {\SiII} and {\SiIII} absorption.  

\begin{figure*} 
\vspace{2.2cm}
\centerline{
\vbox{
\centerline{\hbox{ 
\includegraphics[scale=0.90,angle=00, trim={0 1.5cm 0 2.5cm}]{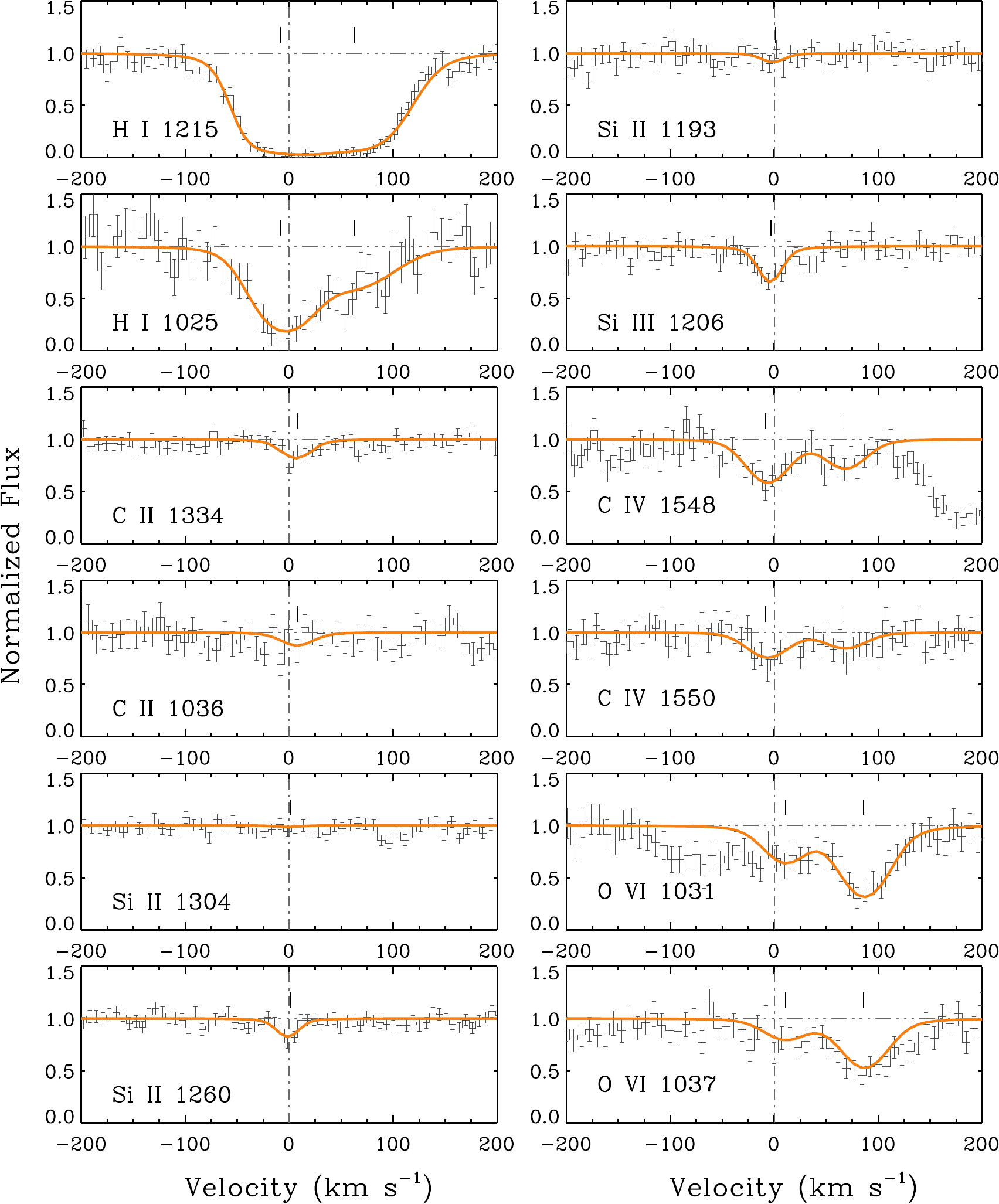} 
}}
}}  
\vspace{1.2cm}
\caption{\small{Voigt profile models superimposed on selected absorption lines in the $z = 0.12122$ absorber towards PG~$1424+240$. The centroids of absorption components are indicated by the vertical tick marks on top of each line. The $1\sigma$ uncertainty in flux is indicated by the error bars. The fit parameters are given in Table 1. The {\OVI}~$1031$ line is contaminated by {\Lyb} from $z = 0.128$ for $-150 < v < 0$~{\kms}. Pixels over this velocity range were de-weighted during the simultaneous fitting of {\OVIdblt} lines.}}  
\label{fig:2} 
\end{figure*}

The {\OVIdblt} also shows absorption in two kinematically distinct clouds. Their centroids derived from profile fitting are offset from {\CII}, {\SiII}, {\CIV} and {\HI} absorption by $\Delta v \sim 20$~{\kms} in both components. (see Table \ref{tab1} and Figure \ref{fig:2}). The difference is only slightly more than the typical residual wavelength calibration errors of $\sim 15$~{\kms} estimated for COS spectra \citep{savage_properties_2014}. Since the {\OVI} absorption spans the same velocity range as {\CIV}, it is possible that the two ions are tracing the same gas phase, or different gas phases that are kinematically intertwined. Such offsets could be due to differences in the velocity fields across the absorbing region, especially when it is extended (\citealt{churchill_direct_2015}). Even if the {\CIV} and {\OVI} are tracing intrinsically different gas phases, it is difficult to segregate their separate contributions in these adjacent absorptions. Thus, while exploring ionization models, we consider the {\CIV} and {\OVI} as tracing the same gas. However, we do note that in many stronger metal line systems, {\CIV} and {\OVI} are inferred to be from different phases of high ionization gas \citep[e.g.,][]{fox_multiphase_2007,lehner_probing_2008}. 

Both members of the {\OVI} doublet suffer varying levels of contamination over some limited velocity ranges of their profiles. The $1031$~{\AA} line is contaminated from $-150$~{\kms} to $-10$~{\kms} by {\Lyb} at $z = 0.128$. From modeling the {\Lya} associated with this interloper, we estimate the extent of contamination (see Figure 1). The residual absorption, after removing the contamination, suggests the presence of {\OVI}~$1031$ accompanying the central {\CIV} component as well, consistent with the {\OVI}~$1037$ at that same velocity. The {\OVI}~$1037$ shows excess absorption over the velocity range $100 \lesssim v \lesssim 150$~{\kms}, which we could not associate with absorption from any known metal line system along this sight line. The wavelength bins affected by contamination were not included during Voigt profile modeling of the {\OVI} doublet lines. The $b$-values of {\CIV} and {\OVI} set an upper limit on temperature of $T \lesssim 2.5 \times 10^5$~K, and $T \lesssim 1.4 \times 10^5$~K for the high ionization gas in the central and offset components respectively, whereas their $1\sigma$ uncertainties suggest that the gas can be hotter with $T \lesssim 5.2 \times 10^5$~K, and $T \lesssim 6.1 \times 10^5$~K respectively.



 Alternatively, one could also use the {\HI}-{\CIV} and {\HI}-{\OVI} to get an ionization-model independent estimate of the temperature. However, one has to be cautious in doing so for a multiphase medium. Absorption line analysis of synthetic spectra derived from simulations with sub-kiloparsec scale resolutions show that metal ions in their high ionization states may not be associated with the observed {\HI} even when they appear as coinciding in velocity space, which in turn adds a systematic bias to the temperature and metallicity estimates (\citealt{oppenheimer_nature_2009}, \citealt{churchill_direct_2015}, \citealt{liang_observing_2018}). Considering this, a more secure upper limit measurement for the temperature would be from the $1\sigma$ $b$-parameter range for the {\CIV} and {\OVI}, as these are ions which tend to probe the same phase. 


\begin{table*} 
\caption{Profile Fit Measurements for the $z = 0.12122$ Absorber towards PG~$1424+240$} 
\begin{center}  
\begin{tabular}{lcccc}
\hline
Line     &	$v$ (\kms)    	   &       log~$[N~(\cmsq)]$	&	$b$(\kms)	&	Component     \\   
\hline
\hline
{\HI}~$1215-1025$	&   $-6~{\pm}~4$		&	$14.81~{\pm}~0.09$	 &	$28~{\pm}~3$	&	Central	\\
			&   $63~{\pm}~6$		&	$14.34~{\pm}~0.05$	 & 	$36~{\pm}~5$	& Offset	\\
\\
{\CIV}~$1548-1550$	&   $-8~{\pm}~3$		&	$13.57~{\pm}~0.06$	 &	$23~{\pm}~4$	&	Central	\\
			&   $67~{\pm}~5$		&	$13.34~{\pm}~0.10$	 &	$24~{\pm}~5$	&	Offset	\\
\\
{\OVI}~$1031-1037$	&   $7~{\pm}~7$			&	$13.84~{\pm}~0.12$	 &	$21~{\pm}~5$	&	Central	\\
			&   $85~{\pm}~4$		&	$14.35~{\pm}~0.10$	 &	$23~{\pm}~4$	&	Offset	\\ \\

{\CII}~$1036-1334$	&   $8~{\pm}~5$		&	$13.27~{\pm}~0.10$	 &	$15~{\pm}~4$	&	Central	\\ \\
{\SiII}~$1193-1260$	&   $1~{\pm}~4$			&	$12.20~{\pm}~0.10$	 &	$10~{\pm}~4$	&	Central	\\ \\
{\SiIII}~$1206$	&   $-3~{\pm}~3$			&	$12.47~{\pm}~0.09$	 &	$10~{\pm}~4$	&	Central	\\

\hline
\hline
\end{tabular}
\label{tab1}
\end{center}
\scriptsize{Comments - The absorption lines were fitted with Voigt profiles using the \citet{fitzpatrick_composition_1997} routine. The models were convolved with the empirically determined COS line-spread functions of \citet{Kriss_2011} at the redshifted wavelength of each line. (a) The {\OVI}~$1031$ line is blended with Ly-$\beta$ absorption from $z = 0.12773$ (confirmed by the presence of corresponding {\Lya}). The contamination has affected the blue part of the {\OVI}~$1031$ feature in the manner shown in Figure 1. Pixels beyond $v = -15$~{\kms} were therefore deweighted during profile fitting. We found that shifting this velocity alters the $b$-parameter significantly. The $b$ value given by the fitting routine does not account for the systematic uncertainty. Taking this contamination also into account, a more realistic measurement should be $b(\OVI) = 21^{+6}_{-10}$~{\kms}. The uncertainty in the corresponding column density is within the $1\sigma$ error given by the profile fitting routine.}
\end{table*}

\begin{table*} 
\caption{Apparent Optical Depth Measurements for the $z = 0.12122$ Absorber towards PG~$1424+240$} 
\begin{center}  
\begin{tabular}{lcccc}
\hline
Line     &	$W_r$~(m{\AA})    	   &       log~$[N_a~(\cmsq)]$	&	$[-v, +v]$~(\kms)	&	Component     \\   
\hline
\hline
{\HI}~$1215$		&   $> 698$		&	$ > 14.5$		& [-100, 200]	&	Central \\
{\HI}~$1025$		&   $> 305$		&	$ > 14.8$		& [-100, 200]	&	Offset \\
\\
{\CIV}~$1548$		&   $109~{\pm}~15$	&	$13.51~{\pm}~0.09$ 	& [-60, 35]	&	Central \\
			&   $57~{\pm}~13$	&	$13.21~{\pm}~0.12$	& [35, 100]	&	Offset \\
\\
{\CIV}~$1550$		&   $45~{\pm}~16$	&	$13.42~{\pm}~0.17$	& [-60, 35]	&	Central \\
			&   $42~{\pm}~13$	&	$13.36~{\pm}~0.17$	& [35, 100]	&	Offset \\
\\
{\OVI}~$1031$		&   $ < 110$		&	$< 14.03$		& [-60, 35]	&	Central \\
			&   $139~{\pm}~14$	&	$14.19~{\pm}~0.06$	& [35, 160]	&	Offset \\
\\
{\OVI}~$1037$		&   $47~{\pm}~12$	&	$13.93~{\pm}~0.10$	& [-60, 35]	&	Central \\
			&   $126~{\pm}~12$	& 	$14.41~{\pm}~0.07$	& [35, 160]	&	Offset \\
\\
{\SiIII}~$1206$		&   $33~{\pm}~10$	& 	$12.24~{\pm}~0.09$	& [-40, 17]	&	Central \\
			&   $< 30$		&	$< 12.2$		& [35, 100]	&	Offset \\	
\\
{\CII}~$1334$		&   $16~{\pm}~5$	&	$12.91~{\pm}~0.16$	& [-20, 15]	&	Central \\
			&   $< 20$		&	$< 13.0$			& [35, 100]	&	Offset \\
\\
{\SiII}~$1260$		&   $15~{\pm}~4$	&	$11.98~{\pm}~0.13$	& [-20, 15]	&	Central \\
			&   $< 18$		&	$< 12.0$		& [35, 100]	&	Offset \\
\\
{\SiII}~$1193$		&   $< 18$		&	$< 12.4$		& [-20, 15]	&	Central \\
			&   $< 27$		&	$< 12.6$		& [35, 100]	&	Offset \\
\\
{\SiIV}~$1403$		&   $< 51$		&	$< 13.1$		& [-60, 100]	&	Central \\
\\
{\NV}~$1238$		&   $< 43$		&	$< 13.3$		& [-60, 100]	&	Central \\
\\
{\AlII}~$1670$		&   $< 267$		& 	$< 12.9$		& [-60, 100]	&	Central \\
\\
{\AlIII}~$1854$		&   $< 153$		&       $< 12.8$		& [-60, 100]	&	Central \\
\\
{\AlIII}~$1862$		&   $< 174$		&	$< 13.3$		& [-60, 100]	&	Central \\	
\hline
\hline
\end{tabular}
\label{tab2}
\end{center}
\scriptsize{Comments - The various columns list the equivalent width of the lines in the rest-frame of the absorber, the integrated apparent column density, and the velocity range for the integration. The upper limits correspond to lines which are not detected at $\geq 3 \sigma$ significance. The lower limits for {\HI} column density indicate that those transitions are heavily saturated. The {\OVI}~$1031$ is contaminated by Ly-$\beta$ absorption from $z = 0.12773$ at the blue end. Hence the measurement here is given as an upper limit. The inconsistent apparent column densities for the {\OVI} component at $v \sim 86$~{\kms} suggests contamination to the {\OVI}~$1037$ which we could not identify with any line associated with known absorbers along this quasar sightline. The plot of apparent column density with velocity (PLOT NOT INCLUDED) clearly shows this excess absorption is in the wings of {\OVI}~$1037$. The aluminum lines are from low $S/N$ STIS E230M grating observations.}
\end{table*}

\section{IONIZATION \& CHEMICAL ABUNDANCES IN THE ABSORBER}\label{section:modelling}

We rely on the measured column densities and their upper limits as constraints to determine the phase structure of the absorber. We consider both photoionization and collisional ionization scenarios. The photoionization models were computed using {\scshape CLOUDY} \citep[version 13.03][]{ferland_2013_2013}, which models the absorbing gas as uniform density plane parallel slabs. The ionizing radiation we adopt is the upgraded extragalactic background radiation model given by \citet{khaire_new_2019}, which incorporates the most recent measurements of the quasar luminosity function \citep{croom_2df-sdss_2009, palanque-delabrouille_luminosity_2013} and star formation rate densities \citep{khaire_star_2015}. The collisional ionization models were based on the calculations of \citet{gnat_time-dependent_2007}. In both cases, the gas is assumed to have solar abundances as given by \citet{asplund_chemical_2009}.

\subsection{Photoionization Modeling}

\subsubsection{The Central Component at $v \sim -6$~{\kms}}

\begin{figure*}[h]
\centerline{
\vbox{
\centerline{\hbox{ 
\includegraphics[scale=0.65,angle=00,angle=90]{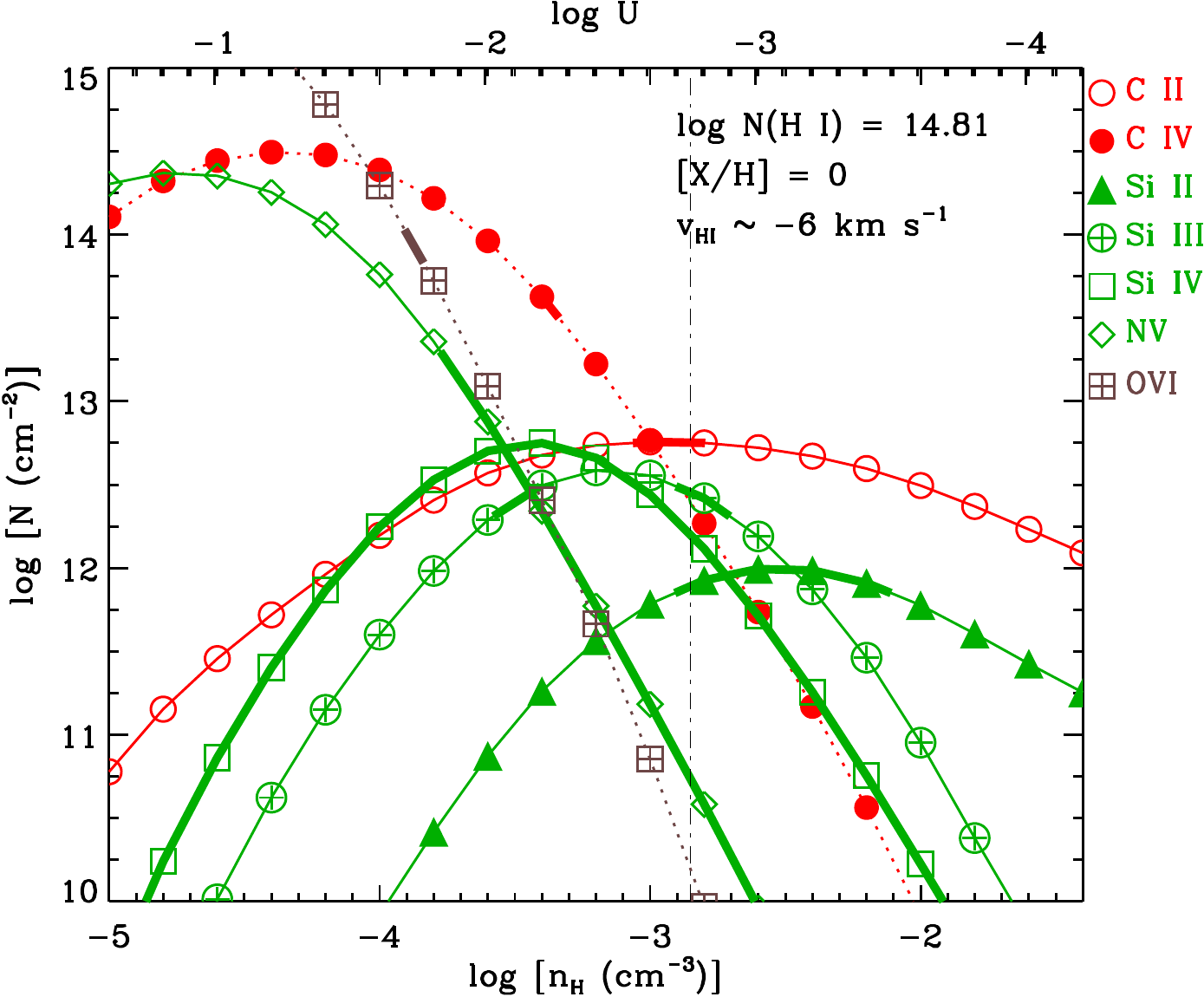} 
\includegraphics[scale=0.65,angle=00,angle=90]{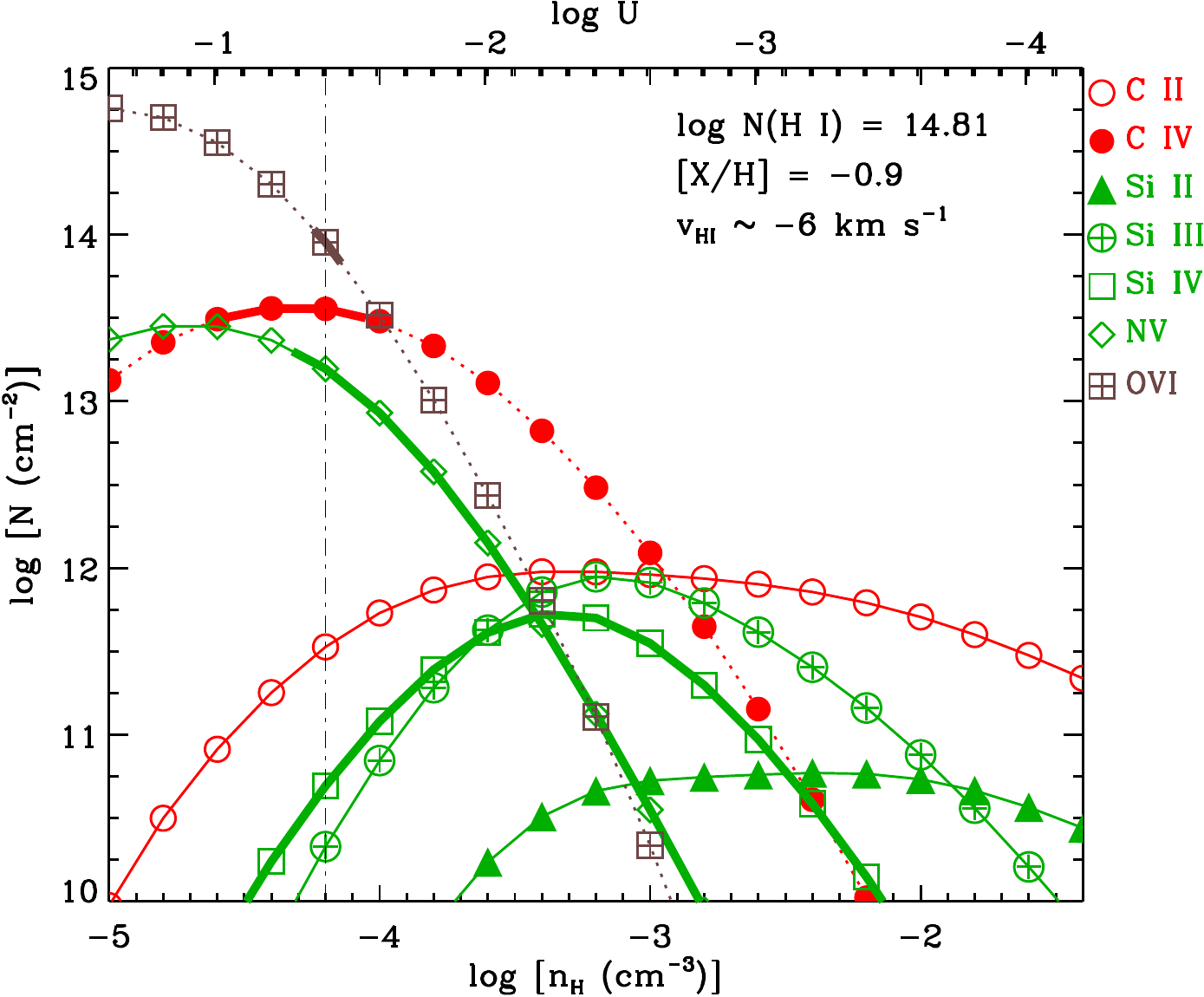} 
}}
}}  
\caption{Photoionization models for the $v_{\HI} = -6$~{\kms} \textit{central} component of the absorber. The bottom X-axis is density, and the top axis is ionization parameter given by $\log U = \log n_{\gamma} - \log n_{\H}$. The curves are column density predictions for the various ions at different gas phase densities. The observed column densities are indicated by thick lines. The single phase solution that simultaneously explains {\CII}, {\SiII}, and {\SiIII} is shown using the \textit{dash-dotted} vertical line in the \textit{left} panel, and the higher ionization solution that explains the {\CIV} and {\OVI} for the central component is shown in the \textit{right} panel.}  
\label{fig:3} 
\end{figure*}

This {\HI} component has detections of low ({\CII}, {\SiII}), intermediate ({\SiIII}) and high ionization species ({\CIV}, {\OVI}). Figure~\ref{fig:3} shows results from photoionization driven equilibrium models for densities in the range $n_{\H} = 10^{-5} - 10^{-1}$~{\cc}, for the observed $\log [N(\HI),~\cmsq] = 14.81$. The models require the carbon abundance to be at least solar to recover the observed {\CII} for any gas density. The observed $N(\SiII)$ places a similar constraint of [Si/H] $\geq 0$.  For [C/H] = [Si/H] $= 0$, the two ions can arise in a single gas phase with $n_{\H} = 1.4 \times 10^{-3}$~{\cc} (ionization parameter, $\log U = \log (n_{\gamma}/n_{\H}) = -2.75$). As shown in Figure~\ref{fig:3}, the same gas phase also reproduces the observed $N(\SiIII)$ and is simultaneously consistent with the non-detections of {\SiIV} and {\NV}. The model also predicts for this phase a total hydrogen column density of $N(\H) \sim 1.6 \times 10^{17}$~{\cmsq}, equilibrium temperature of $T = 9.4 \times 10^3$~K, thermal pressure of $p/K = 13.2$~{\cc}~K, and a line of sight thickness of $L \sim 37$~pc. The temperature is consistent with the narrow {\CII} and {\SiII} profiles. This phase, however, does not explain the {\CIV} and {\OVI} detected at the same velocity. The column density predictions for {\CIV} and {\OVI} are $\sim 1.2$~dex and $\sim 4$~dex lower than the corresponding observed values. A second phase is required to explain the origin of these high ions. 


The presence of multiphase gas at the same velocity has implications for the metallicity estimates.  The $\log N(\HI) = 14.81$ seen for this central component could largely be from the lower ionization {\CII} - {\SiII} gas. The neutral hydrogen accompanying a more highly ionized {\CIV} - {\OVI} phase coinciding in velocity will be comparatively less. In any case, the available data is not adequate to segregate the separate {\HI} contributions from the two gas phases. If indeed $\log~N(\HI)$ is less the the upper limit of $14.81$ in {\CII} - {\SiII} phase, it would lead to a super-solar C and Si abundances in the weakly ionized gas they trace. 

\begin{figure*} 
\centerline{
\vbox{
\centerline{\hbox{ 
\includegraphics[scale=0.85,angle=00,angle=90]{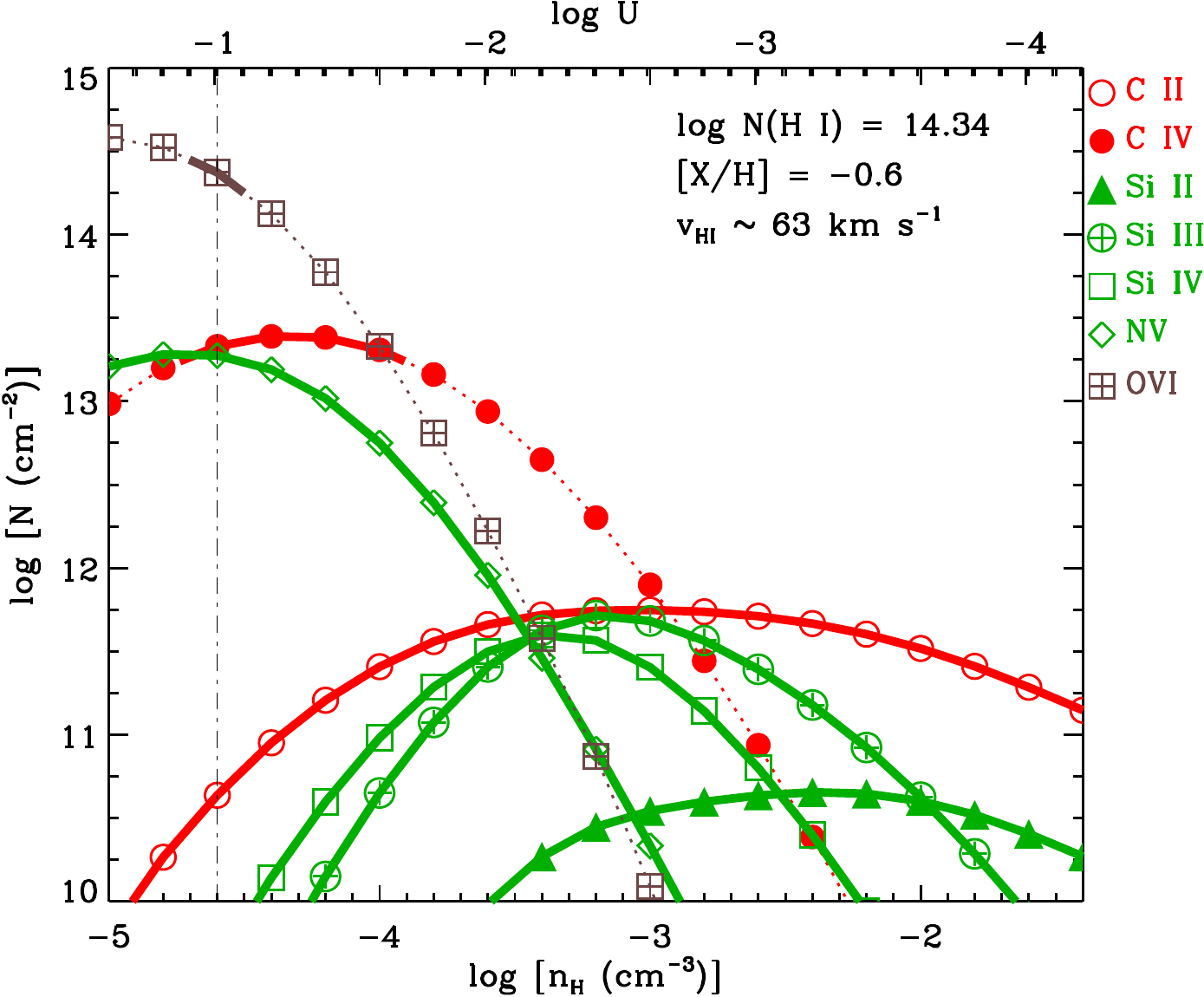} 
}}
}}  
\caption{Photoionization models for the $v_{\HI} = +63$~{\kms} \textit{offset} component of the absorber. The bottom X-axis is density, and the top axis is ionization parameter given by $\log U = \log n_{\gamma} - \log n_{\H}$. The curves are column density predictions for the various ions at different gas phase densities. The observed column densities are indicated by thick lines. The single phase solution that simultaneously explains the {\CIV} and {\OVI} seen at this velocity is represented by the vertical dash-dotted line.}  
\label{fig:4} 
\end{figure*}

For the central component, the {\CIV} and {\OVI} have to come from a separate phase. Under ionization equilibrium, $\log [N(\CIV)/N(\OVI)] = 0.23$ for a density of $n_{\H} \sim 6 \times 10^{-5}$~{\cc}, which is two orders of magnitude lower than the density of the weakly ionized gas (see Figure \ref{fig:4}). It is difficult to determine the exact chemical abundances, total hydrogen column density and size of this high ionization medium because of the uncertain {\HI}. Considering the extreme possibility that most of the {\HI} coinciding with the central component is associated with this {\CIV} and {\OVI}, we obtain a conservative lower bound of $ \gtrsim -0.9$~dex for the carbon and oxygen abundances. For smaller values of {\HI}, higher values of abundances will be required to match the observed {\CIV} and {\OVI} from the same phase. As the abundances approach solar values, the models start producing appreciable amounts of {\CII}. Thus, solar and higher metallicities can be ruled out. The models predict this absorbing medium with $n_{\H} = 6.3 \times 10^{-5}$~{\cc} (assuming a solar [C/O] abundance pattern), with a total hydrogen column density of $\log~N(\H) \leq 19.0$, ~$T \sim 4.1 \times 10^4$~K and a line of sight thickness of~$L \leq 49$~kpc. The temperature is much lower than the upper limit suggested by the $b$-parameters of {\CIV} and {\OVI}. 

\subsubsection{The Offset Component at $v \sim +63$~{\kms}}

At the velocity of this {\HI} component there are no detections of low or intermediate ions, suggesting a predominance of higher ionization conditions in this cloud. The photoionization models are shown in Figure~\ref{fig:5}. The models require [C/H] $\geq -0.6$~dex to generate the observed $N(\CIV)$. Within that abundance constraint, the observed {\CIV} and {\OVI} can come from a single gas phase with $n_{\H} = 2.5 \times 10^{-5}$~{\cc} (assuming solar [C/O] abundance pattern), $\log N(\H) = 18.94$, $T = 3.5 \times 10^4$~K and $L = 112$~kpc. At this density, the model predictions for the low and intermediate ions are consistent with their non-detections. The [C/H] and [O/H] in this cloud are comparable to the abundance limits we derive for both these elements in the higher ionization phase of the central component. We emphasize that the metallicity, total hydrogen column density and absorber size estimates are based on the premise that the {\HI} and the high ions are cospatial and tracing the same phase, an assumption that is contentious (see last paragraph of Sec \ref{section:analysis} and \citealt{churchill_direct_2015}). The photoionization equilibrium temperature predicted by the models is consistent with the upper limit indicated by the combined {\CIV} and {\OVI} line widths.

\subsection{Is Collisional Ionization Significant?}

In many intervening absorbers, {\OVI}, by virtue of its comparatively high ionization potential ($> 114$~eV), is found to have an origin in gas with $T \gtrsim 10^5$~K where the ionization is dictated by collisions between energetic free electrons and metal ions (\citealt{savage_far_2002}, \citealt{howk_global_2002}, \citealt{lehner_connection_2009}, \citealt{narayanan_highly_2010}, \citealt{narayanan_cosmic_2010}, \citealt{savage_cos_2011}, \citealt{narayanan_cosmic_2012}, \citealt{savage_properties_2014}, \citealt{pachat_pair_2016}). The range of temperatures where the fractional abundances of {\CIV} and {\OVI} peak for collisional ionization equilibrium (CIE) conditions falls within the upper limits on the temperatures of the {\CIV} - {\OVI} phase in the central and the offset components, obtained from the different $b$-values of {\CIV} and {\OVI} lines. We therefore consider the prospects of a warm collisionally ionized phase in the absorber. 

\begin{figure*} 
\centerline{
\vbox{
\centerline{\hbox{ 
\includegraphics[scale=0.70,angle=00,angle=90]{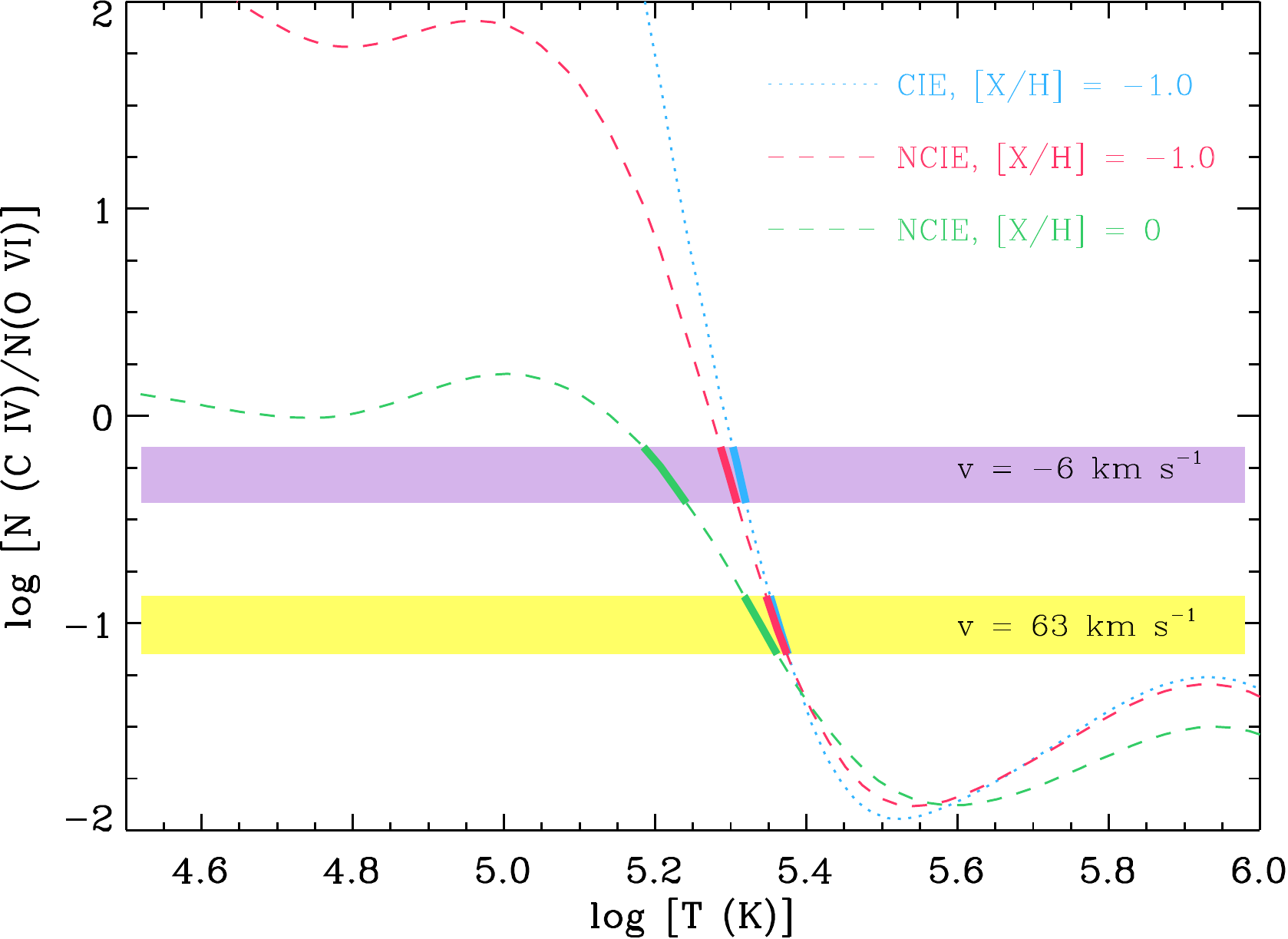} 
}}
}}  
\caption{The CIE (\textit{dotted} line) and non-CIE (\textit{dashed} line) model predictions of the {\CIV} to {\OVI} column density ratio for different gas temperatures. The models are based on Gnat \& Sternberg (2007). The non-CIE predictions are for solar and one-tenth solar metallicities. The \textit{thick} portion in each curve is the $1\sigma$ range of the observed column density ratio. The column density ratio for the $v = -6$~{\kms} (central) and $v = 63$~{\kms} (offset) clouds are indicated by the horizontal shaded regions.}  
\label{fig:5} 
\end{figure*}

In CIE, the ionic column density ratios are dependent only on the equilibrium temperature. The non-CIE scenario is valid for gas that is radiatively cooling from a high temperature, with faster cooling rates for gas of higher metallicity. Under non-equilibrium conditions, {\CIV} and {\OVI} ionization fractions do not decline as rapidly as CIE, for $T < 2 \times 10^5$~K \citep[e.g., see Figure 3 of][]{savage_properties_2014}. Thus, one may expect to see these ions at temperatures of $T \lesssim 10^5$~K. In Figure~\ref{fig:5}  are the CIE and non-CIE predictions for the {\CIV} to {\OVI} column density ratio based on the models of \citet{gnat_time-dependent_2007}.

In the offset component the observed column density ratio of one-tenth between {\CIV} and {\OVI} is reproduced at $T = 2.2 \times 10^5$~K, where both CIE and non-CIE have similar predictions. This estimate complies with the $T \lesssim 6.1 \times 10^5$~K given by the combined {\CIV} and {\OVI} $1\sigma$ line width ranges. The estimate is based on the assumption that the [C/O] relative abundance is solar. Additionally, the $N(\CIV) < N(\OVI)$ seen for this component is true only for $T \gtrsim 2 \times 10^5$~K, which places a constraint on the temperature in this high ionization gas. 

For the central component, the observed $N(\CIV) \lesssim N(\OVI)$ is valid at $T \sim 2 \times 10^5$~K, with the non-CIE models predicting a slightly lower temperature of $T \sim 1.6 \times 10^5$~K for solar metallicity, and [C/O] abundance. It needs to be emphasized that the metallicity in this higher ionization gas is uncertain as the {\HI} at this velocity is more likely to be a part of the cooler photoionized gas traced by the low ions. At the model predicted temperature the {\HI} ionization fraction is only $f(\HI) \sim 3.1 \times 10^{-6}$. The temperature derived from the collisional ionization models for this component are consistent with the upper limit of $T \lesssim 5.2 \times 10^5$~K set by the combined line widths for {\CIV} and {\OVI}.


From the collisional ionization modelling, we conclude that the presence of {\CIV} and {\OVI} can be indicative of warm gas with a temperature of $T \sim (2 - 5.2) \times 10^5$~K for the central component and $T \sim (2 - 6.1) \times 10^5$~K for the offset component, with collisional ionization contributing in very substantial ways in maintaining the high ionization levels. An argument can also be made for the dominance of photoionization in a very diffuse medium ($n_{\H} \sim 10^{-5}$~{\cc}) for the kinematically offset component.

\subsection{Hybrid Models\label{hyb}}

A more vigorous approach to modeling should consider the effects of collisional and photoionization simultaneously. Such hybrid models were also computed to understand the origin of high ions in the central and offset components. Integrating time dependent collisional ionization models (non-CIE) with photoionization is an elaborate exercise beyond the scope of this work. Our hybrid models are restricted to CIE conditions. Hybrid models were computed for temperatures centered around $T \sim 10^5$~K, as suggested by the CIE models and the absorption line widths. Since we do not have a measure on the amount of neutral hydrogen present in the higher ionization phase, the observed {\HI} is taken as an upper limit. For the central and offset components, the hybrid models were able to recover the observed ratio between {\CIV} and {\OVI} column densities for [C/O] of solar, with the ionization dominated by collisions. For the warm gas phase in the central component, hybrid modelling gave us a solution at $T = 1.5 \times 10^5$~K, corresponding to gas densities of $n_{\H} \sim (5-7) \times 10^{-5}$~{\cc}, a total hydrogen column density of $N(\H) \sim 2.5 \times 10^{20}$~{\cmsq}, and oxygen and carbon abundances of [X/H] $> -2.0$. The $N(\H)$ and abundance values are highly conservative limits because of the lack of knowledge about the {\HI} associated with the warm gas phase. For the offset warm gas phase we obtain a hybrid solution at $T = 2 \times 10^5$~K, giving similar values of $n_{\H} \sim (8-10) \times 10^{-5}$~{\cc}, $N(\H) \sim 1.1 \times 10^{20}$~{\cmsq}, oxygen and carbon abundances of [X/H] $> -1.5$ and a line of sight thickness of $\sim 373$~kpc.\\

We note that the path length obtained for the high ionization gas through the different models are notably large. For example, while hybrid models for the highly ionized  $v \sim +63$~{\kms} offset component yield a line-of-sight thickness of $L \sim 373$~kpc, photoionization modelling of the same cloud suggests a size of $L \sim 112$~kpc. These size scales, though large, are consistent with the $\sim 280$~kpc transverse length scales and $L \sim 600-800$~kpc line-of-sight thicknesses reported by \citet{muzahid_probing_2014} for a large-scale {\OVI}-absorbing structure in the multiphase CGM of a 1.2L* galaxy at $z = 0.2$, observed using a pair of quasar sightlines, with one of the sightlines also showing a weak {\MgII} analogue absorber.

\section{GALAXIES NEAR THE ABSORBER}\label{section:galaxies}

An important step towards deciphering the origin of the absorber is to extract information about galaxies located in its vicinity. In this section, we describe our search for galaxies neighbouring the z = 0.1212 absorber using the database of the Sloan Digital Sky Survey, which covers the PG~$1424+240$ sightline in its footprint. In the SDSS DR15 spectroscopic database, we queried for galaxies with spectroscopic redshifts that are within a projected separation of $\rho = 1$~Mpc from the PG1424+240 quasar sightline and $|\Delta v| = 750$~{\kms} of the absorber redshift (z = 0.1212). These constraints were adopted from the typical radius and velocity dispersion of galaxy clusters as reported in \citealt{bahcall_large-scale_1988}. This exercise revealed that there are two galaxies in the search field. Figure~\ref{fig:galimage}  shows a mosaic image of $7^{\prime} \times 7^{\prime}$ centered on the background quasar PG~$1424+240$, with the two nearby galaxies labelled. The nearest of these two galaxies is located within 200 kpc of the absorber, and was identified by \citep{muzahid_cos-weak:_2018} as well. This $1.4L^*$ galaxy, at a redshift of z $=$ 0.12118, also has a very small velocity offset of 11 km/s with respect to the absorber redshift, suggesting a possible association between the two. 

\begin{figure*}
\centering
\begin{subfigure}{42pc}
  \centering
  \includegraphics[scale=0.26]{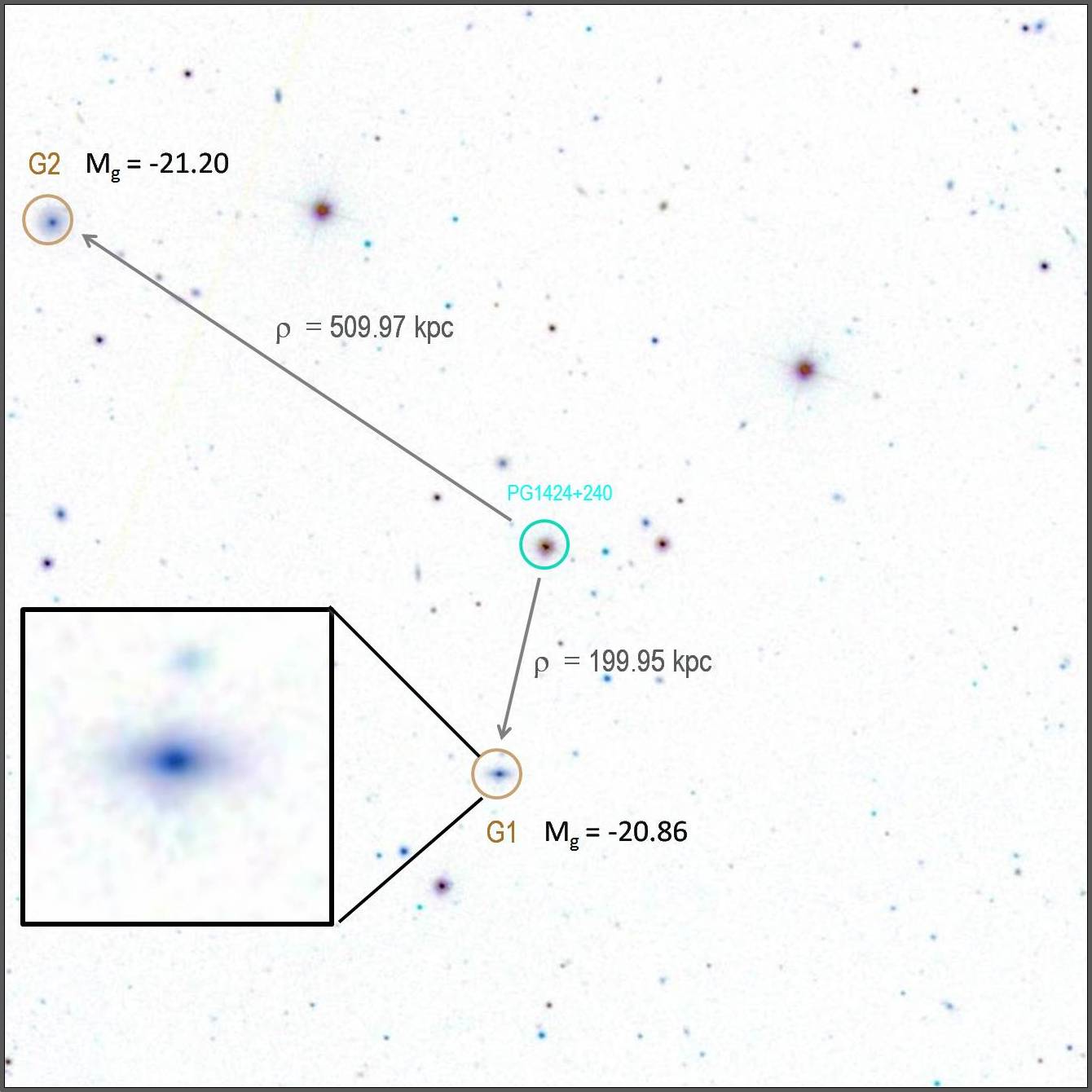}
\end{subfigure}
\vspace{0.1in}
\begin{subfigure}{42pc}
  \centering
  \includegraphics[scale=0.45,angle=90]{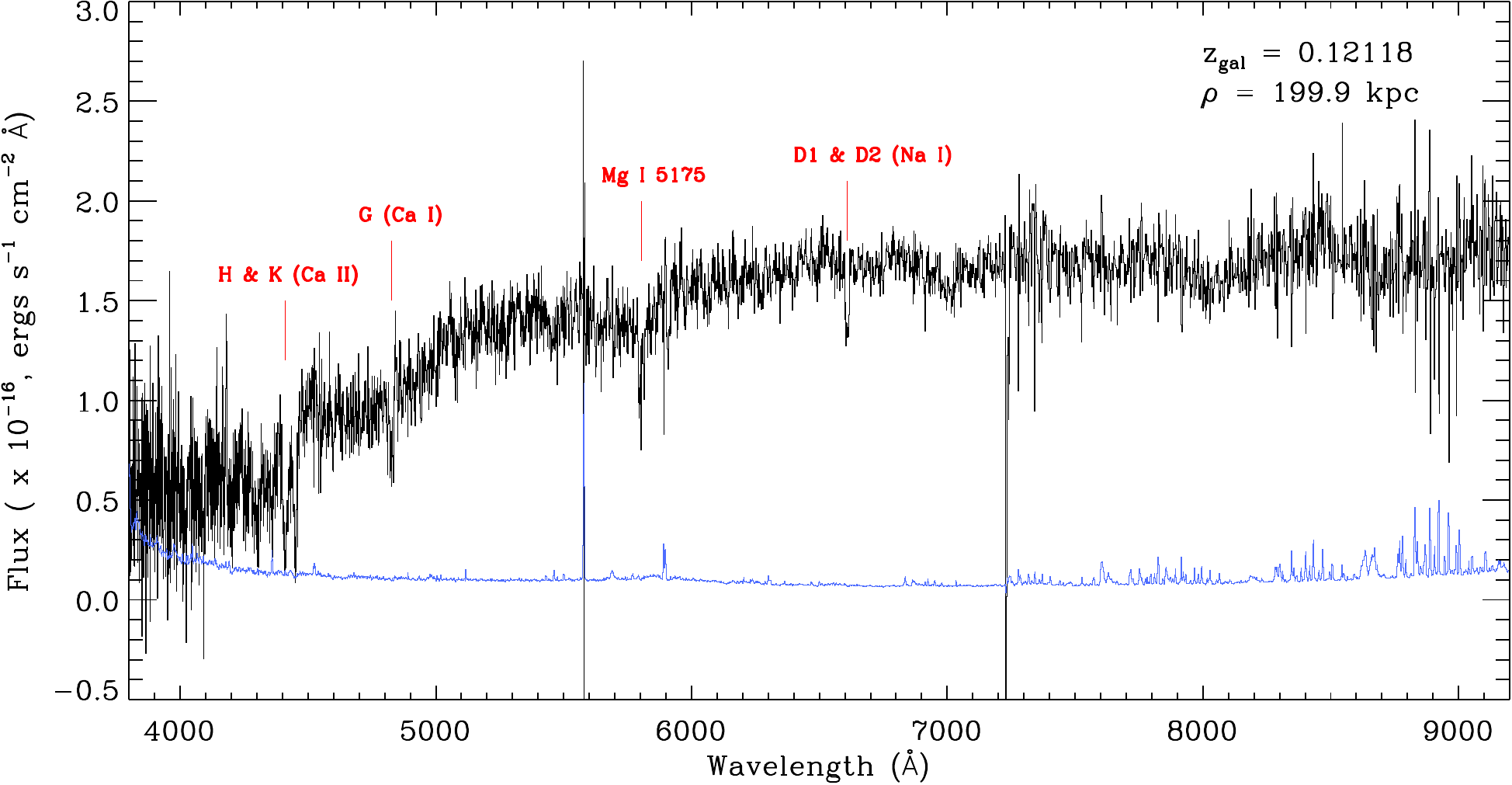}
\end{subfigure}
\caption{An SDSS $g,~r,~i$ three color composite mosaic image of the field centered on the quasar PG~$1424+240$ (\textit{right} panel). The image is $7^{\prime}~\times~7^{\prime}$ across. The galaxies labeled G1 and G2 are nearest to the $z=0.12122$ absorber in projected separation. The inset shows a zoomed-in image of galaxy G1 which is within $\rho = 200$~kpc from the line of sight and has a velocity separation $<$ 50 km/s with the z = 0.1212 absorber, indicating a possible association of the absorber with this galaxy. The low dispersion ($R \sim 2000$) SDSS spectrum of G1 is shown in the \textit{bottom} panel. }  
\label{fig:galimage} 
\end{figure*}   

\begin{figure*}
\centering
 \vspace{1.5cm}
\includegraphics[trim=3cm 0cm 2.5cm 3cm,scale=0.45]{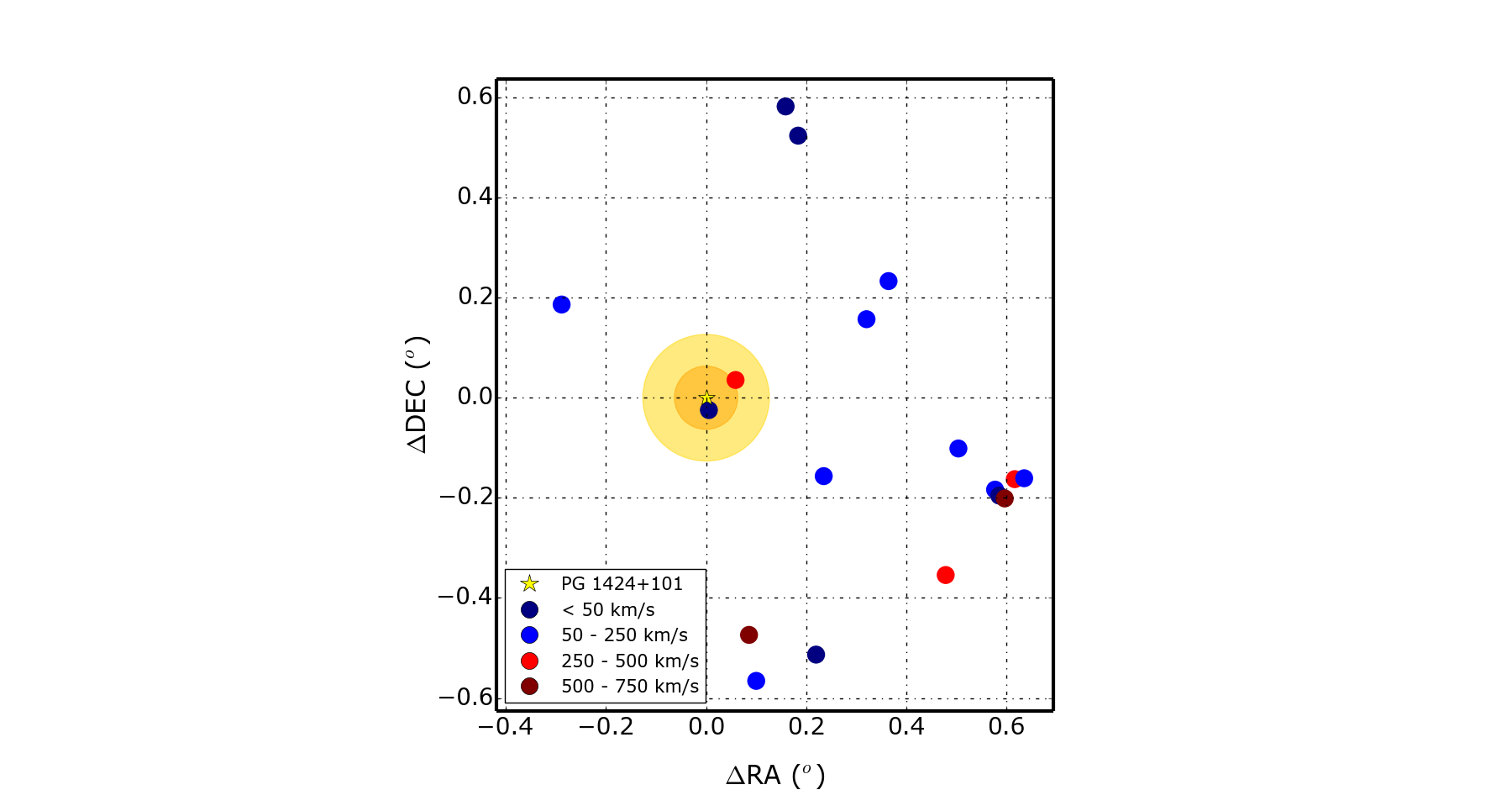}
\caption{Plot shows the spatial distribution of galaxies detected in the SDSS within 750 km s$^{-1}$ of z = 0.1212, and within a radius of 5 Mpc from the RA-Dec of the PG1424+240 quasar sightline, which is denoted by the yellow star. We find 18 galaxies within the search field, hinting that the absorber resides in a galaxy overdensity region. The orange circle represents the region in the sky with a radius of 200 kpc from the sightline, while the light yellow circle denotes the corresponding 1 Mpc radius. The red and blue circles denote the galaxies, with the colour coding corresponding to the velocity separation of the galaxy from the absorber. Out of the 18 galaxies, there are two that lie within 1 Mpc of the quasar sightline, one of which is within 200 kpc and has a velocity separation $<$ 50 km/s from the z = 0.1212 absorber.}
\label{fig:galplot} 
\end{figure*}

In order to look for galaxy associations at scales larger than typical cluster sizes and check whether our absorber is in a galaxy overdensity region, we relaxed the SDSS database search criteria to encompass a projected separation of $\rho = 5$~Mpc and a velocity offset of $|\Delta v| = 750$~{\kms}.  This search identified a total of 18 galaxies with spectroscopic redshifts that are within the field of search. The particulars of these galaxies are listed in Table \ref{tab3}. In Figure \ref{fig:galplot}, the spatial distribution of these 18 galaxies has been plotted with the axes denoting the RA and Dec offsets with respect to the position of the quasar. As evident from the table, all these galaxies have a Luminosity greater than the L* at this epoch. The detection of a significant number of >L* galaxies in the relatively narrow search field hints that our absorber is residing in a galaxy overdensity region. 


\begin{table*} 
\caption{\textsc{Galaxies in the vicinity of the absorber}}
\begin{center}
\small
\begin{tabular}{lccrcccc}
\hline
R.A.  &  Dec.   &   $z_{gal}$   &  $\Delta v$ (\kms)  &   $\rho$  (Mpc)  &  $g$ (mag)  &   M$_g$ & ($L/L^*$)$_g$ \\
\hline
$216.75722$  &	$23.77526$  &   $0.12118$  &  	$-11~{\pm}~11$	&	 $0.2$		&	$18.55$	& $-20.57$  &  	$1.4$ \\
$216.81052$  &  $23.83541$  &   $0.11951$  &	$-457~{\pm}~10$ &	 $0.5$   	&	$17.86$	& $-21.20$  &	$2.6$ \\
$216.98676$  &  $23.64333$  &   $0.12033$  &	$-237~{\pm}~10$ &	 $2.1$   	&	$17.01$	& $-22.10$  &	$5.8$ \\
$216.46294$  &  $23.98611$  &   $0.12214$  &    $247~{\pm}~14$ 	&	 $2.6$		&	$18.12$	& $-21.03$  &	$2.2$ \\
$217.07227$  &  $23.95682$  &   $0.12083$  &    $-105~{\pm}~13$ &	 $2.6$		&   	$18.10$	& $-21.01$  & 	$2.1$ \\
$217.11630$  &  $24.03294$  &   $0.12081$  &    $-110~{\pm}~10$ &	 $3.2$   	&	$17.74$	& $-21.34$  &	$2.9$ \\
$217.25592$  &	$23.69855$  &	$0.12212$  &	$239~{\pm}~11$	&	 $3.7$		&   	$17.63$	& $-21.48$  &	$3.3$ \\
$216.83746$  &	$23.32621$  &   $0.12351$  &    $613~{\pm}~14$ 	&	 $3.8$   	&	$18.67$	& $-20.46$  &	$1.3$ \\
$216.93561$  &  $24.32347$  &   $0.12110$  &    $-33~{\pm}~14$  &        $4.3$  	&    	$18.67$	& $-20.45$  &	$1.3$ \\
$216.97148$  &	$23.28683$  &	$0.12128$  &    $16~{\pm}~10$	&	 $4.4$   	&	$16.98$	& $-21.98$  &	$5.2$ \\
$217.32919$  &  $23.61636$  &   $0.12071$  &	$-138~{\pm}~13$ &	 $4.4$   	&	$18.17$	& $-20.95$  &	$2.0$ \\
$217.23053$  &  $23.44577$  &   $0.12003$  &    $-319~{\pm}~10$ &        $4.5$  	&    	$17.55$	& $-21.55$  &	$3.5$ \\
$217.33759$  &  $23.60450$  &   $0.12129$  &    $19~{\pm}~7$    &        $4.5$  	&	$17.02$	& $-22.11$  &	$5.9$ \\
$216.85185$  &	$23.23434$  & 	$0.12146$  &	$64~{\pm}~5$	&	 $4.5$   	&	$18.31$	& $-20.58$  &	$1.4$ \\
$217.34856$  &  $23.59874$  &   $0.11917$  &    $-548~{\pm}~11$ &        $4.6$   	&	$18.42$	& $-20.64$  &	$1.5$ \\
$217.36858$  &  $23.63731$  &   $0.12230$  &    $290~{\pm}~14$  &        $4.7$  	&    	$18.34$	& $-20.82$  &	$1.8$ \\
$216.91058$  &  $24.38187$  &   $0.12119$  &    $-7~{\pm}~10$   &        $4.7$  	&    	$17.37$	& $-21.65$  &	$3.8$ \\
$217.38739$  &  $23.63910$  &   $0.12081$  &    $-110~{\pm}~13$ &        $4.8$  	&    	$18.31$	& $-20.81$  &	$1.8$ \\
\\
\hline
\end{tabular}
\label{tab3}
\end{center}
\scriptsize{Comments. - Galaxies within a projected separation of $\sim 5$~Mpc and $|\Delta v| = 750$~{\kms} of the absorber. The $z$ values are SDSS spectroscopic redshifts. $\Delta v$ is the systemic velocities of the galaxies with respect to the absorber. The error in velocity separation comes from the uncertainty in the spectroscopic redshift. The projected separation $\rho$ was calculated from the angular separation assuming a $\Lambda$CDM universe with parameters of $H_0 = 69.6$~{\kms}~Mpc$^{-1}$, $\Omega_m = 0.286$, $\Omega_{\Lambda} = 0.714$ \citet{bennett_1_2014}. While determining the absolute magnitudes, appropriate K-corrections were applied using the analytical expression given by \citet{chilingarian_analytical_2010}. The Schecter absolute magnitude $M^*_g = -20.18$ for $z = 0.1$ was taken from\citet{ilbert_vimos-vlt_2005}}
\end{table*}

\begin{figure*}
\centering
\begin{subfigure}
  \centering
  \includegraphics[scale=0.32]{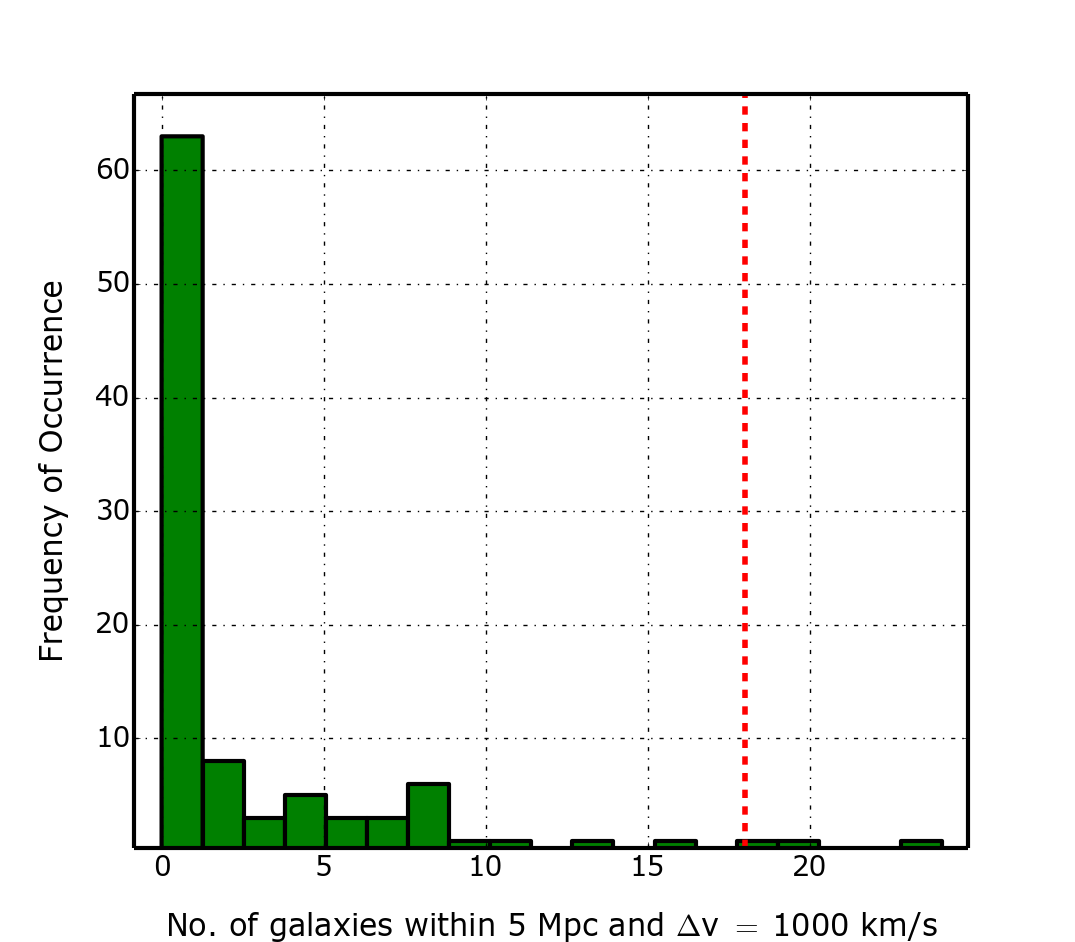}
\end{subfigure}%
\begin{subfigure}
  \centering
  \includegraphics[scale=0.32]{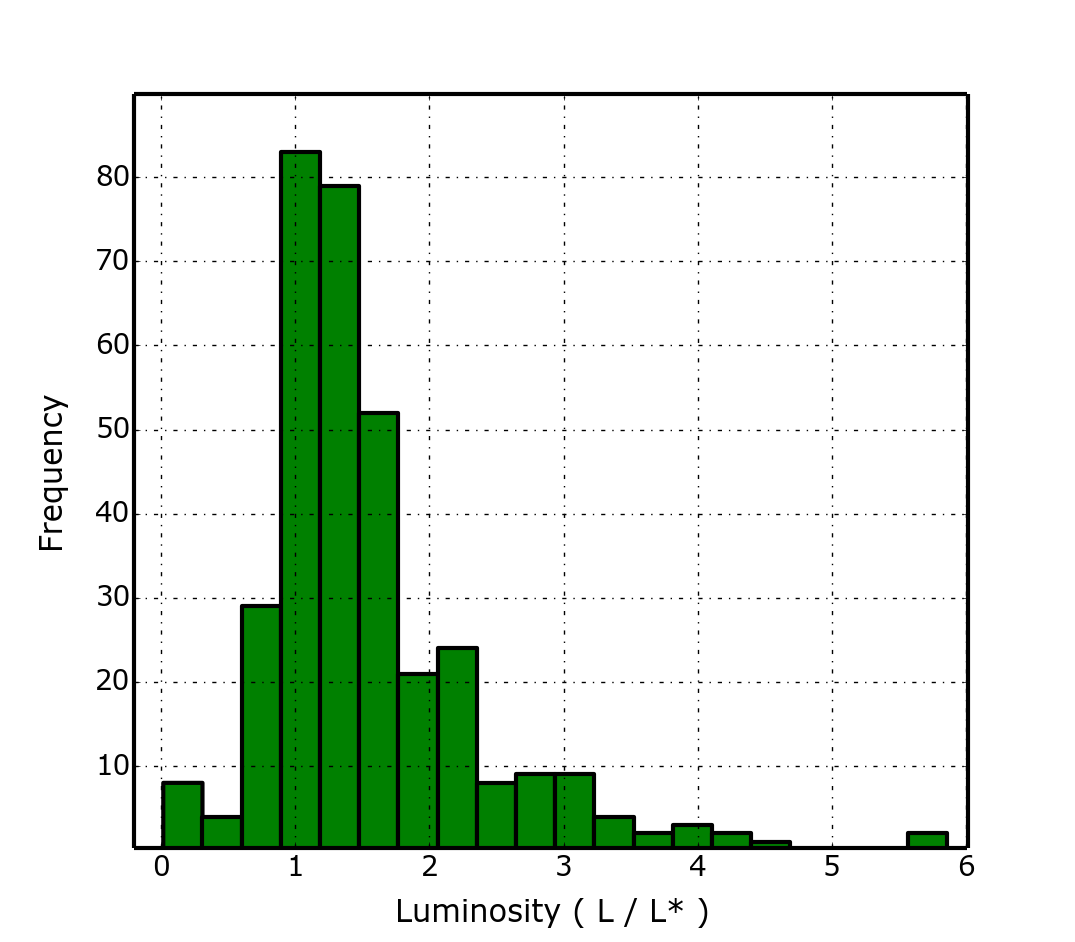}
\end{subfigure}
\caption{The \textit{left} panel shows the frequency distribution of galaxies within 1000 km s$^{-1}$ of $z = 0.1212$, and within a radius of 5 Mpc from the RA-Dec of 100 locations obtained by random sampling of the SDSS footprint. The \textit{dashed red} line corresponds to the number of galaxies that are identified in SDSS for the same search parameters around our absorber. It is evident that 70\% of the fields reveal less than 3 galaxies, and the probability of random coincidence of finding 18 or more galaxies is extremely low ($<$ 5 \%). Our absorber indeed seems to be in a galaxy overdense region. The galaxies found through this random sampling exercise were analyzed to construct the luminosity distribution, which is shown in the \textit{right} panel. The distribution clearly shows that SDSS has poor completeness at $z = 0.1212$ for sub-L* luminosities. This is consistent with the $> L^*$ luminosity derived for all the galaxies in the field around our absorber.}  
\label{fig:MCMC} 
\end{figure*}   

We performed a Monte Carlo sampling of the SDSS DR15 spectroscopic database, to explore whether the the number of bright galaxies detected in this field is higher than what is likely to be observed in a similar random patch of the sky at this redshift. One hundred regions covered in the SDSS footprint were selected at random, and were probed for SDSS-detected galaxies within 1000 km s$^{-1}$ of our absorber redshift of z = 0.1212, and within a 5 Mpc search radius.  Figure \ref{fig:MCMC} (left panel) shows the frequency distribution of the galaxies found in these random search fields. The histogram shows a steep fall in the distribution for higher numbers of galaxies in the search regions, indicating that regions are more likely to have fewer galaxies. In 70\% of the cases, we find less than 3 galaxies in the search field whereas, we find 18 galaxies in the neighbourhood of our absorber and this number is marked by the vertical red dashed line in the frequency distribution plot. From the Monte Carlo analysis, we find that that the probability of randomly detecting 18 or more galaxies in the sampled volume is exceptionally low ($<$ 5 \%), providing a strong evidence in favour of our absorber indeed tracing a galaxy overdensity region, such as a large scale structure of associated galaxies, with a size much larger than typical clusters. 

We also need to verify whether the 18 galaxies detected by SDSS are in fact the only galaxies in the field around our absorber. The fact that all these galaxies have $>$ L* luminosity indicates that the SDSS is incomplete at sub-L* luminosities, and that we are unable to detect fainter galaxies in the region. To investigate this, we obtained the luminosity distribution of the galaxies detected in the Monte Carlo samples, as shown in the right panel of Figure \ref{fig:MCMC}. The distribution shows a distinct fall-off in the number of galaxies having a luminosity less than 1L*, revealing the poor completeness of the SDSS for sub-L* galaxies at z $=$ 0.1212. Hence, SDSS is only sampling  the brightest galaxies in the region around our absorber, and even more galaxies such as dwarfs or other sub-L* galaxies are likely to be present in the vicinity of the absorber, that are too faint to be detected in the SDSS.

The inferred overabundance of galaxies near the z $=$ 0.1212 absorber suggests that the absorption is most likely tracing gas in a large group/cluster environment, with a probable origin in the extended halos of one of the galaxies belonging to that region. The length scales at which we are probing are much greater than those of typical galaxy cluster and group environments, suggesting that the overdensity we are probing is part of a much larger-scale galaxy association. We estimate the average velocity of the SDSS-detected group of galaxies within this association as $269$~{\kms}, with respect to the absorber . The  z $=$ 0.12118 galaxy identified by SDSS as closest to the absorber is a $1.4L^*$ galaxy at an impact parameter of $200$~kpc, and a velocity offset, $\Delta v = -11~{\pm}~11$~{\kms} with reference to the absorber redshift. Using the scaling relation $R_{vir} = 250 (L/L^*)^{0.2}$~kpc (\citealt{prochaska_probing_2011}), we estimate a halo virial radius of $R_{vir} = 267$~kpc for this galaxy. The luminosity scaling relationship of \citet{stocke_absorption-line_2014}, log R$_{vir}$ $=$ 2.257 + 0.318C + 0.018$C^{2}$ - 0.005$C^{3}$ , where C = log (L/L*),  also yields a comparable halo radius of $R_{vir} = 201$~kpc for this galaxy, meaning that the absorber is seemingly located at the fringes of the virial halo of the galaxy. This galaxy has an extended morphology, with an absorption line dominated spectrum (shown in Figure \label{fig:galimage}), suggesting little ongoing star formation. The $u - r = 2.88$ color of the galaxy is also consistent with with being late-type (E/S0, or Sa, \citealt{strateva_color_2001}). From the absence of emission-line flux at H-$\alpha$, the star formation rate in the galaxy is constrained to $\lesssim 0.05$~M$_{\odot}~\mathrm{yr}^{-1}$ (\citealt{kennicutt_global_1998}). The presence of a luminous galaxy not too far in projected separation and velocity from the absorber, and the intriguing similarities between the absorber and multiphase Galactic HVCs, motivate us to consider the absorber as tracing high velocity gas in the extended halo of this $z= 0.12118$ galaxy. Alternate scenarios for the origin are also discussed in the final section. 

\section{SUMMARY \& DISCUSSION}\label{section:summary}

We have carried out absorption line analysis and ionization modelling of a multiphase system at  $z = 0.12122$ towards the blazar PG~$1424+240$, using high resolution HST/COS spectroscopic data. The key results are:

\begin{enumerate}

\item The absorber contains low ionization {\CII} 1334 and {\SiII} 1260 lines, and an intermediate ionization {\SiIII} 1206 line, coinciding in velocity with comparatively stronger detections of higher ionization {\CIVdblt} and  {\OVIdblt} lines. The {\CII} 1334 and {\SiII} 1260 lines are very weak and their measured rest-frame equivalent widths (16 m{\AA} and 15 m{\AA}, respectively) are consistent with this absorber being a weak {\MgII} analog. 

\item The {\CIVdblt} and {\OVIdblt} have at least two kinematically distinct components at $v \sim -6$~{\kms} and $v \sim +63$~{\kms}. Absorption in {\HI} also spans this same velocity range. The weak {\CII}, {\SiII} and {\SiIII} are only detected in the central $v \sim -6$~{\kms} component. 

\item While no single photoionized phase can simultaneously explain the low and high ions in the central component, a multiphase solution is possible with the low ions arising from solar metallicity gas with $n_{\H} \sim 10^{-3}$~{\cc}, and $T \sim 10^4$~K, with a small line of sight thickness of $\sim 37$~parsec. The high metallicity and compact size of the cloud are typical of weak {\MgII} absorbers. The strong {\CIV} and {\OVI} could arise in a warmer ($T \sim 10^5$~K) phase of the gas. These collisional ionization models together with the combined line widths of {\CIV} and {\OVI} limit the temperature in this phase to $T \sim (2 - 5.2) \times 10^5$~K. Alternatively, a  photoionization origin in a low density medium with $n_{\H} \sim 10^{-5}$~{\cc}, $T \sim 10^4$~K, $L \leq 49$~kpc, and [C/H] $\sim $ [O/H] $\gtrsim -0.9$~dex, cannot be ruled out. The multiphase nature of this cloud (parsec scale compact and cool low ionization gas coexisting with diffuse high ionization gas) has strong similarities with Galactic HVCs. 

\item For the {\CIV} - {\OVI} component offset in velocity, photoionization yields a feasible single-phase solution with $n_{\H} \sim 10^{-5}$~{\cc}, $T \sim 3.5 \times 10^4$~K, $\log~N(\H)$ = 18.94, along with carbon and oxygen abundances of $\sim -0.6$~dex, assuming that the {\HI} at this velocity is from the same phase as the high ions. The temperature prediction from the model is consistent with the upper limit obtained from the combined $b$ parameters of $\CIV$ and $\OVI$. Collisional ionization conditions are also possible with temperatures of $T \sim (2 - 6.1) \times 10^5$~K.

\item SDSS shows 18 luminous ($> L^*$) galaxies within $\rho = 5$~Mpc and $750$~{\kms} of systemic velocity from the absorber, indicating that the absorber is probing gas that is part of a galaxy rich environment such as a cluster or a massive group. The nearest galaxy to the absorber is at $\rho = 200$~kpc and $\Delta v = -11~{\pm}~11$~{\kms}. 

\end{enumerate}

The impact parameter of $\rho < R_{vir}$ and the close match in redshift with the nearest galaxy enhances the likelihood of the absorber being located within the CGM of this bright ($1.4L^*$) galaxy. Such an origin is consistent with previous propositions of weak low-ionization absorbers as extragalactic analogs of the Milky Way HVCs (\citealt{lynch_survey_2006}, \citealt{narayanan_survey_2007}, \citealt{mshar_kinematic_2007}, \citealt{narayanan_chemical_2008}, \citealt{richter_population_2009}). The solar metallicity for the low ionization dense gas is within the range found for cool circumgalactic clouds around luminous ($\geq L^*$) galaxies even for projected distances of $R_{vir}$ (e.g., \citealt{prochaska_cos-halos_2017}). Though our knowledge of the association of weak {\MgII} absorbers with galaxies is limited (\citealt{churchill_mgii_2005}, \citealt{chen_what_2010}), both {\CIV} and {\OVI} are known to be strongly clustered around $\sim L^*$ galaxies (\citealt{chen_origin_2001}, \citealt{tumlinson_multiphase_2011}, \citealt{prochaska_probing_2011}, \citealt{werk_cos-halos_2013},  \citealt{stocke_characterizing_2013}, \citealt{savage_properties_2014}, \citealt{churchill_direct_2015}, \citealt{kacprzak_azimuthal_2015}, \citealt{werk_cos-halos_2016}), with a higher covering fraction seen for more massive galaxies ($M_* \gtrsim 10^{10}$~M$_{\odot}$) than low mass ones (e.g., \citealt{burchett_deep_2016}). 

While we consider the possibility of the absorber being an HVC analog in the CGM of the 1.4L* galaxy, we note that the impact parameter of 200 kpc places the cloud much farther out in the halo compared to the highly ionized population of Galactic HVCs, majority of which are within 50 kpc from the Galactic center (\citealt{wakker_distances_2001}, \citealt{thom_galactic_2006}, \citealt{wakker_distances_2007}, \citealt{thom_accurate_2008}, \citealt{wakker_distances_2008}). The subgroup of Compact HVCs (CHVCs), located much farther, at distances of several hundred kiloparsecs, are postulated to have a Local Group origin (\citealt{blitz_high-velocity_1999}, \citealt{braun_kinematic_1999}, \citealt{sembach_highly_2003}). Moreover, the low and high ionization properties in these Galactic compact clouds are not sufficiently well understood to draw a direct comparison with our absorber. Instead, insights can be drawn from a comparison of the absorber with M31's CGM, which is well studied in both absorption and emission. Much of the high neutral column density gas surrounding M31 detected via 21 cm emission are within the inner halo at $\sim$ 0.2 R$_{vir}$ (\citealt{thilker_continuing_2004}, \citealt{westmeier_westerbork_2005}, \citealt{westmeier_relics_2007}, \citealt{westmeier_relics_2008}), with R$_{vir}$ $\sim$ 300. However, a more extended study of the M31 CGM by \citet{lehner_evidence_2015} with better column density sensitivity over multiple background lines of sights has found the presence of metals tracing multiphase diffuse gas out to R$_{vir}$ $\sim$ 300 kpc and possibly even beyond, with the ionization levels of the CGM increasing with projected separation. The 1.4L* galaxy at 200 kpc of projected separation from our absorber likely occupies a more massive and extended halo, suggesting that the absorber could, in fact, be tracing its multiphase CGM. The small column density ratio between low and high ions is also consistent with the dominance of higher ionization gas at large projected distances, as seen for the M31 CGM

\citet{schaye_large_2007} had proposed that metals transported through the CGM by supernova driven winds or similar large-scale gas outflows could be confined as compact clouds poorly mixed with the ambient medium. These patchy clouds will be transient as they undergo free expansion due to pressure imbalance with the surrounding gas. The thermal gas pressure in the low ionization phase of our absorber ($p/k \sim 13.2$~{\cc}~K) is an order of magnitude higher than the pressure in the higher ionization gas ($p/k \sim 2.46$~{\cc}~K), if it is photoionized. The high-metallicity, higher density gas traced by {\CII} and {\SiII} could be a patchy transient cloud in a more diffuse higher ionization medium traced by {\CIV} and {\OVI}. As the ionization models indicate, the {\CIV} and {\OVI} coinciding with the weak low ionization gas can be collisionally ionized, in which case they could be transition temperature plasma at the conductive interface or turbulent mixing layers between the cool ($T \sim 10^4$~K) low ionization cloud and the $T \sim 10^6$~K ambient corona of the luminous galaxy or a hot intragroup medium. For several extragalactic absorbers, collisional ionization in such interface layers has been the favored mechanism for the origin of {\OVI} (\citealt{narayanan_highly_2010}, \citealt{savage_o_2010}, \citealt{savage_cos_2011}, \citealt{tumlinson_multiphase_2011}, \citealt{tripp_hidden_2011}, \citealt{pointon_impact_2017}, \citealt{stocke_absorption-line_2014}, \citealt{stocke_warm_2017}). \\

The same is also postulated for the origin of the highly ionized phase in Galactic HVCs (\citealt{borkowski_radiative_1990}, \citealt{slavin_turbulent_1993}, \citealt{breitschwerdt_delayed_1994}, \citealt{murphy_far_2000}, \citealt{sembach_far_2000}, \citealt{sembach_highly_2003}, \citealt{collins_highly_2005}, \citealt{fox_multiphase_2005}, \citealt{fox_study_2008}). Considering hybrid models (refer Section \ref{hyb}) for the higher ionization phase of the central component, if the higher ionization gas is indeed at an equilibrium temperature of $1.5 \times 10^5$~K, then the two phases are more closer to being in pressure balance ($p/k \sim 9$~{\cc}~K). 

Using the length scale and temperature from the photoionization model for the low ionization gas, we compute its free expansion time-scale to be $t \sim R/c_s \sim 2.5$~million years \footnote{Here $R$ is the size of the absorbing cloud, $c_s = \sqrt{\gamma kT/\mu m_H}$ is the sound speed, $\gamma = 5/3$ and the average mass per particle $\mu = 0.62$ for a predominantly ionized gas with mass fractions of H, He and heavier elements as 0.70, 0.28 and 0.02 respectively.}, which is much smaller compared to the Hubble time. This may imply a low likelihood of catching such transient structures in random lines of sight studies. However, weak low ionization clouds have a comoving number density that could be six orders of magnitude\footnote{This estimate assumes spherical geometry for the absorbing clouds} higher compared to bright L > 0.1L* galaxies (\citealt{rigby_population_2002}, \citealt{schaye_large_2007}). Though transient in nature, the far more numerous presence of these gas clouds in regions clustered around galaxies compensates for their short lifetimes, yielding a higher rate of incidence in absorption sightline surveys (\citealt{rigby_population_2002}, \citealt{lynch_physical_2007}, \citealt{narayanan_chemical_2008}).  

If the cloud is embedded in the CGM of the 1.4L* galaxy, a useful exercise is to compare the model-derived gas pressure values with the expected ambient pressure of the CGM at an impact parameter of 200 kpc from this galaxy. In the case of the Milky Way, \citet{bouma_population_2019} provide a scaling relation between CGM gas pressure and Galactocentric radius, based on the coronal gas model by \citet{miller_constraining_2015}. Adopting this Galactic pressure profile for studying the CGM of the 1.4 L* galaxy, we compute a circumgalactic gas pressure of $p/k \sim 7.65$~{\cc}~K at  $\rho \sim 200$~kpc, where our absorber resides. This pressure falls in between the $p/k \sim 13.2$~{\cc}~K obtained for the low ionization phase and the $p/k \sim 2.46$~{\cc}~K for the higher ionization phase, derived via photoionization modeling for the central component. The computed CGM pressure value is also comparable to the $p/k \sim 9$~{\cc}~K derived through hybrid models for the higher phase of the central component.      

Explanations tending towards both photoionization and collisional ionization can be made for the {\CIV} and {\OVI} in the offset cloud. In fact, the origin of {\CIV} and {\OVI} in intervening absorbers has always been ambiguous. The ionization fraction of both ions peak at temperatures slightly above $T \sim 10^5$~K under collisional ionization conditions (\citealt{gnat_time-dependent_2007}). At the same time photoionization can also contribute in very substantial ways in low density gas (\citealt{tripp_high-resolution_2008}, \citealt{hussain_implications_2017}), and is likely to dominate in high-redshift absorbers due to a higher intensity of the UV extragalactic background radiation. The theoretical models of \citealt{oppenheimer_nature_2009} categorized strong {\OVI} absorbers ($N \gtrsim 10^{14.5}$~{\cmsq}) as tracing collisional ionization in warm gas, whereas weaker {\OVI} lines were more in agreement with a photoionized scenario (also see \citealt{cen_revealing_2001}, \citealt{fang_probing_2001}, \citealt{chen_x-ray_2003}, \citealt{cen_where_2006}). The mass of the dark matter halo of the host galaxy has a strong influence on the ionization mechanism in circumgalactic clouds. Massive $L^*$ halos possess virial temperatures sufficient to produce {\CIV} and {\OVI} through collisional ionization (e.g. \citealt{prochaska_probing_2011}, \citealt{pointon_impact_2017}, \citealt{mason_inferences_2019}), whereas in sub-$L^*$ halos (M $\lesssim$ 10$^{11}M_{\odot}$) it is governed by photoionization (\citealt{churchill_ionization_2014}, \citealt{oppenheimer_bimodality_2016}). In evaluating the present absorber, it may be apt to also think of the absorption arising from sub-$L^*$ halos, rather than the more massive and luminous $z = 0.12118$ galaxy identified by SDSS as closest to the absorber, which we discuss next.  

While the association with a metal rich outflow in the CGM of the $z = 0.12118$ galaxy is a possibility, there are alternatives that can be considered. For example, this galaxy nearest in impact parameter is part of a massive group or cluster of several very bright galaxies (see Section \ref{section:galaxies} and Figure \ref{fig:galplot}). The line-of-sight could very well be probing sub-$L^*$ galaxies undetected by SDSS and closer in separation. The luminosity distribution function of SDSS galaxies at the redshift of the absorber exposes the significant incompleteness of the SDSS galaxy spectroscopic database at fainter luminosities (see Figure \ref{fig:MCMC}).  Weak low-ionization absorbers have previously been interpreted to be associated with dwarf galaxies. The redshift evolution of these absorbers peak at $z \sim 1.2$ coinciding with the peak in global star formation rate in dwarf galaxies, hinting at a possible causal relation between dwarf galaxies and at least some fraction of weak {\MgII} absorbers (\citealt{lynch_survey_2006}, \citealt{narayanan_survey_2007}, \citealt{lynch_physical_2007}). More conclusive evidence is presented in other works, where weak low-ionization absorbers were found to be tracing supernova driven outflows in close proximity with starburst and post-starburst dwarf galaxies (\citealt{zonak_absorption_2004}, \citealt{stocke_discovery_2004}, \citealt{keeney_discovery_2006}). 

Simulations show strong outflows from supernovae creating a widespread distribution of metals around star-forming dwarf galaxies ($M_h \lesssim 10^{10}$~M$_{\odot}$, $R_{vir} \lesssim 50$~kpc) because of their shallow gravitational potential (\citealt{shen_baryon_2014}, \citealt{oppenheimer_bimodality_2016}, \citealt{muratov_metal_2017}, \citealt{christensen_tracing_2018}). On the other hand, absorption line studies probing the CGM of sub-$L^*$ and dwarf galaxies have consistently found a low covering fraction of metals, especially in their lower ionization states, even for impact parameters of $\sim 0.5 R_{vir}$, compared to massive galaxies (\citealt{prochaska_probing_2011}, \citealt{bordoloi_cos-dwarfs_2014}, \citealt{liang_mining_2014}, \citealt{burchett_deep_2016}, \citealt{johnson_extent_2017}). If this is the case, then our line of sight has to be probing the CGM at close projected distances, much less than half the virial radius from the central sub-$L^*$ galaxy. 

At the same time, it is also important to factor in the environment in which these dwarf galaxies reside. \citet{johnson_extent_2017} in their study of CGM absorbers have only considered isolated field dwarf galaxies. The weak absorber we study, instead, resides in a galaxy rich environment where tidal forces from massive galaxies and/or ram pressure induced by their hot circumgalactic corona can displace interstellar gas from dwarf satellite galaxies significantly enhancing the covering fraction of cool low ionization gas around them. Such gaseous tidal streams are known to be a source of high velocity gas in the CGM of Milky Way and M31 (\citealt{putman_magellanic_2003}, \citealt{thilker_continuing_2004}, \citealt{fox_multiphase_2005}, \citealt{fox_exploring_2010}, \citealt{besla_simulations_2010}, \citealt{besla_role_2012}, \citealt{donghia_magellanic_2016}). Along similar lines, in the multiphase CGM simulations of \citet{oppenheimer_multiphase_2018}, the presence of neighboring galaxies within $300$~kpc, as in a group environment, is found to increase the incidence of low ionization gas at distances further out in the CGM ($> 100$~kpc). Thus, the weak absorber originating in an undiscovered dwarf galaxy in an overdense environment remains an intriguing possibility. The {\CIV} and {\OVI} coincident in velocity with the low ions could be interface gas between a cooler cloud and a hotter medium, whereas the {\CIV} and {\OVI} offset in velocity could be tracing the ambient photoionized gas in a kiloparsec scale region within the CGM. The two {\CIV} clouds have very similar conditions suggesting similar origins.  

An alternative is for the weak absorber to be tracing intra-group gas common to the large scale galaxy environment, rather than the gaseous halo of an individual galaxy. Numerous previous absorption line studies of galaxy dense environments have found the presence of high metallicity gas in the intragroup/intracluster space (\citealt{rasmussen_gas_stripping_2006}, \citealt{McConnachie_2007}, \citealt{freeland_intergalactic_2011}, \citealt{zavala_removal_2012}, \citealt{boselli_quenching_2016},  \citealt{gavazzi_ubiquitous_2018}, \citealt{burchett_warm_2018}, \citealt{manuwal_civ_2019}, \citealt{pradeep_detection_2019}) beyond the $R_{vir}$ of the nearest bright galaxies. The chemical enrichment of such gas is understood as due to the displacement of metals from galaxies via physical processes like feedback from supernovae, AGNs, tidal interactions between galaxies or stripping of interstellar gas due to ram pressures as galaxies move through the hot group/cluster medium. Such mechanisms have been proposed by previous studies to explain the origin of other weak {\MgII} absorbers (\citealt{narayanan_chemical_2008}, \citealt{muzahid_cos-weak:_2018}).

Finding weak low ionization absorbers with additional coverage of {\HI}, {\CIV} and {\OVI} in the combined COS and STIS archive is potentially an important means for identifying metal-rich HVC analogues surrounding other galaxies. Even as large surveys capture the broad statistics of absorber populations, it is also important to assess individual absorption systems in detail, establish the chemical abundances of elements in them and the ionization phases they trace, all of which are vital for understanding their origin. The measurements in this paper demonstrate the need to do this with high $S/N$ spectral data complemented by deep galaxy observations that go down to faint luminosities in the extended volume encompassing the absorber. 

\section*{Acknowledgements}

We thank the anonymous referee for a useful and constructive review of the manuscript. Support for this work was provided by SERB through grant number EMR/2017/002531 from the Department of Science \& Technology, Government of India. This work is based on observations made with the NASA/ESA Hubble Space Telescope, support for which was given by NASA through grant HST GO-14655 from the Space Telescope Science Institute. STScI is operated by the Association of Universities for Research in Astronomy, Inc. under NASA contract NAS 5-26555. VK was supported by NASA through grant number HST-AR-15032.002-A from the STScI. This research has made use of the HSLA database, developed and maintained at STScI, Baltimore, USA, and also the Sloan Digital Sky Survey (SDSS) data base. Funding for the Sloan Digital Sky Survey IV has been provided by the Alfred P. Sloan Foundation, the U.S. Department of Energy Office of Science, and the Participating Institutions. JC acknowledges support from NSF AST-1517816.




\bibliographystyle{mnras}

\bibliography{HVC_weak} 




\appendix


\bsp	
\label{lastpage}
\end{document}